\documentstyle[12pt,aaspp4]{article}

\slugcomment{To appear in Astrophys. J. 535 (2000)}

\lefthead{Matsubara} \righthead{
CORRELATION FUNCTION IN REDSHIFT SPACE}

\begin{document}
\newcommand{\gsim}{\mbox{\raisebox{-1.0ex}{$\stackrel{\textstyle >}
{\textstyle \sim}$ }}}
\newcommand{\lsim}{\mbox{\raisebox{-1.0ex}{$\stackrel{\textstyle <}
{\textstyle \sim}$ }}}

\newcommand{\gtsima}{$\; \buildrel > \over \sim \;$}
\newcommand{\ltsima}{$\; \buildrel < \over \sim \;$}
\newcommand{\simgt}{\lower.5ex\hbox{\gtsima}}
\newcommand{\simlt}{\lower.5ex\hbox{\ltsima}}
\newcommand{\himpc}{{\hbox {$h^{-1}$}{\rm Mpc}} }
\newcommand{\bfk}{{\mbox{\boldmath $k$}}}
\newcommand{\bfq}{{\mbox{\boldmath $q$}}}
\newcommand{\bfr}{{\mbox{\boldmath $r$}}}
\newcommand{\bfx}{{\mbox{\boldmath $x$}}}
\newcommand{\bfy}{{\mbox{\boldmath $y$}}}
\newcommand{\bfz}{{\mbox{\boldmath $z$}}}
\newcommand{\bfv}{{\mbox{\boldmath $v$}}}
\newcommand{\sbfk}{{\mbox{\scriptsize\boldmath $k$}}}
\newcommand{\sbfx}{{\mbox{\scriptsize\boldmath $x$}}}
\newcommand{\bfpsi}{{\mbox{\boldmath $\psi$}}}
\newcommand{\bfPsi}{{\mbox{\boldmath $\Psi$}}}
\newcommand{\mPsi}{{\mit\Psi}}
\newcommand{\tth}{\widetilde{\theta}}
\newcommand{\thh}{\theta_{\rm h}}
\newcommand{\tdel}{\widetilde{\delta}}
\def\pp{\par\parshape 2 0truecm 15.5truecm 1truecm 14.5truecm\noindent}
\renewcommand{\theequation}{\mbox{\rm
{\arabic{section}.\arabic{equation}}}} 


\title{THE CORRELATION FUNCTION IN REDSHIFT SPACE: GENERAL FORMULA
WITH WIDE-ANGLE EFFECTS AND COSMOLOGICAL DISTORTIONS}

\author{Takahiko Matsubara\footnote{Department of Physics, The
        University of Tokyo, Hongo 7-3-1, Tokyo 113-0033, Japan; {\it
        and} Research Center for the Early Universe, Faculty of
        Science, The University of Tokyo, Tokyo 113-0033, Japan.}}

\affil{Department of Physics and Astronomy, 
        The Johns Hopkins University,
        3400 N.Charles Street, Baltimore, MD 21218
}

\begin{abstract}

A general formula for the correlation function in redshift space is
derived in linear theory. The formula simultaneously includes
wide-angle effects and cosmological distortions. The formula is
applicable to any pair with arbitrary angle $\theta$ between lines of
sight, and arbitrary redshifts, $z_1$, $z_2$, which are not
necessarily small. The effects of the spatial curvature both on
geometry and on fluctuation spectrum are properly taken into account,
and thus our formula holds in a Friedman-Lema\^{\i}tre universe with
arbitrary cosmological parameters $\Omega_0$ and $\lambda_0$. We
illustrate the pattern of the resulting correlation function with
several models, and also show that validity region of the conventional
distant observer approximation is $\theta \le 10^\circ$.

\end{abstract}


\keywords{cosmology: theory --- galaxies: distances and redshifts ---
quasars: general --- large-scale structure of universe --- methods:
statistical}

\newpage

\section{INTRODUCTION}
\label{sec1}
\setcounter{equation}{0}

Ever since the pioneering work of Totsuji \& Kihara (1969) and Peebles
(1974), the two-point correlation function of galaxies has been one of
the most fundamental tools in analyzing the large-scale structure of
the universe. Recently, prominent advances in galaxy redshift surveys
have been taking place (for review, see Strauss 1999), and upcoming
large-scale galaxy and QSO surveys, notably the Two-Degree Field
Survey (2dF; Colless 1998; Folkes 1999), and the Sloan Digital Sky
Survey (SDSS; Gunn \& Weinberg 1995; Margon 1998), will provide
three-dimensional, large-scale redshift maps of galaxies and QSO's.
The two-point correlation function will play one of the central role
in the analysis of such large-scale redshift maps.

In redshift surveys, the distances to objects are measured by
recession velocities, and thus the distribution of objects in redshift
space is not identical to that in real space. The clustering pattern
is distorted by peculiar velocity fields (Kaiser 1987), and, for high
redshift objects, also by cosmological warp of real space on a
light-cone (Alcock \& Paczy\'nski 1979). Such effects are called as
{\em redshift distortions}. These effects on the linear power spectrum
and on the linear two-point correlation function have been
investigated by many authors (for review, see Hamilton 1998)

Most of the work is for redshift distortions in a nearby universe and
assumes $z \ll 1$. Kaiser (1987), in his seminal paper, derived the
linear redshift distortions of the power spectrum for a nearby
universe, employing the distant-observer approximation, which assumes
the scales of fluctuations of interest are much smaller than the
distances to the objects. Generally, the redshift distortions by
peculiar velocities are along the line of sight. This radial nature of
distortion introduces a statistical inhomogeneity into the redshift
space. In the distant-observer approximation, such inhomogeneity is
neglected and the statistical homogeneity is recovered, while the
anisotropy is introduced, instead. The Fourier spectrum has maximal
advantage when the statistical homogeneity does exist, thus, for this
reason, Kaiser's formula has a very simple form. Hamilton (1992)
transformed Kaiser's formula of the power spectrum to the formula of
the two-point correlation function, using the Legendre expansion (see
also Lilje \& Efstathiou 1989; McGill 1990).

Despite the simple form of the formula, the distant-observer
approximation is not desirable in the sense that we cannot utilize the
whole information of the survey within this approximation. The
wide-angle effect, which is the contribution of an angle $\theta$
between lines of sight of two objects to the correlation function,
affects the analysis of the redshift maps when we intend to use the
whole data of the survey. Therefore, the analysis of wide-angle effect
on the correlation function in linear theory (Hamilton \& Culhane
1996; Zaroubi \& Hoffmann 1996; Szalay, Matsubara \& Landy 1998;
Bharadwaj 1999), as well as the spherical harmonic analysis (Fisher,
Scharf \& Lahav 1994; Heavens \& Taylor 1995), are of great
importance.

All the above studies are for a nearby universe, and assume $z \ll 1$.
This condition is not appropriate for modern galaxy redshift surveys
($z \sim 0.2$) or QSO redshift surveys ($z \sim 2$). The evolution of
clustering and the nonlinearity in the redshift-distance relation
introduce {\em the cosmological redshift distortion} on the
correlation function. Ballinger, Peacock \& Heavens (1996) and
Matsubara \& Suto (1996) explored this effect in power spectrum and
correlation function, respectively, both employing the
distant-observer approximation (see also Nakamura, Matsubara \& Suto
1997; de Laix \& Starkman 1997; Popowski et al. 1998; Nishioka \&
Yamamoto 1999; Nair 1999). The more the depth of the redshift surveys
is increasing, the more the cosmological redshift distortion becomes
important.

So far the previous formulas for two-point correlation function in
redshift space are restricted either to nearby universe, or to
distant-observer approximation. The purpose of the present paper is to
derive the unified formula for the linear redshift distortions of the
correlation function, fully taking into account the wide-angle effects
and cosmological distortions, simultaneously. In this way, we do not
have to think about which formula we should use depending on the value
of $z$ and $\theta$. In addition, the effects of the spatial curvature
are also included in our formulation and our formula applies to
Friedman-Lema\^{\i}tre universe with arbitrary cosmological parameters
$\Omega_0$ and $\lambda_0$. Our formula turns out to correctly
reproduce the known results if we take appropriate limits of the
formula.

In \S\ref{sec2}, the redshift-space distortions of density
fluctuations toward radial directions are derived on a light-cone,
which include the evolutionary effects. In \S\ref{sec3}, the formula
for correlation function in redshift space is derived separately for
open, flat, and closed models. Demonstrations of redshift distortions
in several cases and the validity region of the conventional distant
observer approximation are given in \S\ref{sec4}. The conclusions are
summarized in \S\ref{sec5}.

\section{LINEAR REDSHIFT-SPACE DISTORTIONS ON A LIGHT-CONE}
\label{sec2}
\setcounter{equation}{0}

\subsection{Radial Distortion of Linear Density Field on a Light-cone}
\label{sec2.1}

Throughout the present paper, we assume the
Friedman-Lema\^{\i}tre-Robertson-Walker (FLRW) metric as the
background space-time of the universe. According to the sign of the
spatial curvature $K$, the metric is given by
\begin{eqnarray}
   ds^2 = - c^2 dt^2 + \left\{
   \begin{array}{ll}
      \displaystyle
      \frac{a^2(t)}{-K}
      \left[
         d\chi^2 + \sinh^2\chi(d\theta^2 + \sin^2\theta d\phi^2)
      \right],
      & (K<0) \\
      \displaystyle
      a^2(t)
      \left[
         d\chi^2 + \chi^2 (d\theta^2 + \sin^2\theta d\phi^2)
      \right],
      & (K=0) \\
      \displaystyle
      \frac{a^2(t)}{K}
      \left[
         d\chi^2 + \sin^2\chi(d\theta^2 + \sin^2\theta d\phi^2)
      \right],
      & (K>0)
   \end{array}\right.
\label{eq1}
\end{eqnarray}
where $\theta$ and $\phi$ are usual angular coordinates, and $\chi$ is
a radial comoving coordinate. In the following, we choose a unit
system, $c = 1$, and $H_0 = 1$, where $H_0 = 100h{\rm km/s/Mpc}$ is
the Hubble's constant, so that comoving distances are measured in
units of $c H_0^{-1} = 2997.9 \himpc$. In this unit system, the
spatial curvature $K$ is then given by
\begin{eqnarray}
&&
   K = \Omega_0 + \lambda_0 - 1,
\label{eq2}
\end{eqnarray}
where the scale factor at present is normalized as $a_0 = 1$. {}From
the FLRW metric, we define the three dimensional metric $\gamma_{ij}$
as
\begin{eqnarray}
   ds^2 = - dt^2 + a^2 \gamma_{ij}dx^i dx^j.
\label{eq5.1}
\end{eqnarray}
That is, the tensor $\gamma_{ij}$ is the metric for a three-space of
uniform spatial curvature $K$. In the following, vectors with Latin
indices represent the three-vector in the three-space specified by the
metric $\gamma_{ij}$.

The comoving distance $x$ at redshift $z$ is given by
\begin{eqnarray}
   x(z) = \int_0^{z} \frac{dz'}{H(z')},
\label{eq3}
\end{eqnarray}
where $H(z)$ is the Hubble parameter at (comoving) redshift $z$:
\begin{eqnarray}
   H(z) = 
  \sqrt{(1 + z)^3 \Omega_0 + 
  (1 + z)^2 (1 - \Omega_0 - \lambda_0) + \lambda_0}.
\label{eq4}
\end{eqnarray}
The radial comoving coordinate $\chi$ and the comoving distance $x$
are related by
\begin{eqnarray}
   \chi = \left\{
   \begin{array}{ll}
      |K|^{1/2} x(z), & (K \ne 0) \\
      x(z). & (K = 0)
   \end{array}
   \right.
\label{eq5}
\end{eqnarray}

Provided that a cosmological model is fixed, the comoving distance
$\chi$, and the redshift $z$ are related through equation (\ref{eq5})
on the observable past light-cone, and we interchangeably use these
variables in the following. Therefore, we use the redshift $z$ as an
alternative spatial coordinate. On a light cone surface, it is also
considered as an alternative of the time variable, $t$.

In this paper, the light ray is assumed to be transmitted on an
unperturbed metric for simplicity. This point is discussed later.
Then, in redshift space, the angular coordinates ($\theta, \phi$) are
common to those in real space. Only the radial distances are distorted
in redshift space. Let us consider an object located at comoving
coordinates $(t, x^i)$ which has the 4-velocity $u^\mu$, normalized as
$u^\mu u_\mu = -1$ as usual. In terms of the three-dimensional
peculiar velocity $v^i$, this 4-velocity is given by
\begin{eqnarray}
   u^{\mu}  =
   \frac{\left(1, v^i\right)}{\sqrt{1 - a^2 v^i v_i}}.
\label{eq6}
\end{eqnarray}
The wave 4-vector, $k^\mu$, satisfying the null geodesic equations,
$k^\mu k_\mu = 0$ and $k^\mu_{;\nu} k^\nu = 0$, is given by
\begin{eqnarray}
   k^\mu \propto 
   \left(a^{-1}, -a^{-2} n^i\right),
   \qquad
   k_\mu \propto \left(-a^{-1}, -n_i\right),
\label{eq6.5}
\end{eqnarray}
where $n^i$ is a three dimensional normal vector, which represents the
line of sight, and $n_i = \gamma_{ij} n^j$. The frequencies of the
light at the source and at the observer are given by $\nu_1 = (k_\mu
u^\mu)|_t$, and $\nu_0 = (k_\mu u^\mu)|_{t_0}$, respectively. Assuming
the peculiar velocities are non-relativistic, the redshift $z_{\rm
obs}$ the observer actually observes is given by
\begin{eqnarray}
   1 + z_{\rm obs} =
   \frac{\nu_1}{\nu_0} = 
   (1 + z)(1 + W - W_0)
\label{eq7}
\end{eqnarray}
where $W = a n_i v^i$ and $W_0 = n_i v_0^{\,i}$ are the line-of-sight
components of peculiar velocities of the source and of the observer.
The peculiar velocity $W$ is evaluated on a light cone, and $W_0$ can
be estimated from the value of the dipole anisotropy of cosmic
microwave background radiation.

For convenience, we define the redshift-space physical comoving
distance as $s(z) = x(z_{\rm obs})$ where $z_{\rm obs}$ is given by
the equation (\ref{eq7}), and the function $x(z)$ is formally the same
as equation (\ref{eq3}) by definition. Explicitly, $s(z)$ is defined
by
\begin{eqnarray}
&&
   s(z) = 
   \int_0^{z + (1+z)(W - W_0)}
   \frac{dz'}{H(z')}.
\label{eq7.1}
\end{eqnarray}
In other words, the redshift-space physical comoving distance $s$
defined by equation (\ref{eq7.1}) is the apparent physical comoving
distance of an object in redshift space, which is originally at
redshift $z$ in real space, and is shifted by its own peculiar
velocity. In a limit $z_{\rm obs} \rightarrow 0$, equation
(\ref{eq7.1}) reduces to the usual relation for nearby universe
(Kaiser 1987), $s = z_{\rm obs}/H_0 = x + (W - W_0)/H_0$. From the
redshift-space physical comoving distance $s$, we also define the
redshift-space analog of comoving coordinate, $\chi_s$ as $\chi_s(z) =
\chi(z_{\rm obs})$, or equivalently,
\begin{eqnarray}
   \chi_s = \left\{
   \begin{array}{ll}
      |K|^{1/2} s(z), & (K \ne 0) \\
      s(z). & (K = 0)
   \end{array}
   \right.
\label{eq7.2}
\end{eqnarray}

The difference between the number density of observed objects in real
space, $n^{\rm (r)}(\chi,\theta,\phi,t)$ and that in redshift space,
$n^{\rm (s)}(\chi_s,\theta,\phi)$, are related by the number
conservation:
\begin{eqnarray}
&&
   n^{\rm (s)}(\chi_s,\theta,\phi) \sinh^2\chi_s d\chi_s d\Omega =
   n^{\rm (r)}(\chi,\theta,\phi) \sinh^2\chi d\chi d\Omega,
   \qquad (K < 0)
\nonumber\\
&&
   n^{\rm (s)}(\chi_s,\theta,\phi) \chi_s^{\,2} d\chi_s d\Omega =
   n^{\rm (r)}(\chi,\theta,\phi) \chi^2 d\chi d\Omega,
   \qquad\qquad\qquad\;\; (K = 0)
\label{eq7.5}\\
&&
   n^{\rm (s)}(\chi_s,\theta,\phi) \sin^2\chi_s d\chi_s d\Omega =
   n^{\rm (r)}(\chi,\theta,\phi) \sin^2\chi d\chi d\Omega,
   \qquad\quad\; (K > 0)
\nonumber
\end{eqnarray}
where $d\Omega = \sin\theta d\theta d\phi$, and $n^{\rm (r)}$ is
evaluated on a light cone. Therefore,
\begin{eqnarray}
   n^{\rm (s)}(\chi_s,\theta,\phi) = 
   n^{\rm (r)}(\chi,\theta,\phi) \times
   \left\{
   \begin{array}{ll}
      \displaystyle
      \left(
         \frac{\sinh^2\chi_s}{\sinh^2\chi}
         \frac{\partial\chi_s}{\partial\chi}
      \right)^{-1},
      & (K < 0) \\
      \displaystyle
      \left(
         \frac{\chi_s^{\,2}}{\chi^2}
         \frac{\partial\chi_s}{\partial\chi}
      \right)^{-1},
      & (K = 0) \\
      \displaystyle
      \left(
         \frac{\sin^2\chi_s}{\sin^2\chi}
         \frac{\partial\chi_s}{\partial\chi}
      \right)^{-1}.
      & (K > 0)
   \end{array}
   \right.
\label{eq8}
\end{eqnarray}
The number density in real space is given by the underlying number
density of objects $\rho$ multiplied by the selection function $\Phi$:
\begin{eqnarray}
   n^{\rm (r)}(\chi,\theta,\phi) = 
   \Phi(\chi,\theta,\phi) \rho(\chi,\theta,\phi)
\label{eq9}
\end{eqnarray}
where we allow the direction dependence of the selection function.

We assume the linear theory of density fluctuation throughout this
paper and we only consider the first order in perturbation with
respect to the variables, $W$, $W_0$, and $\delta = \rho/\bar{\rho} -
1$. Then the relation between $\chi_s$ and $\chi$ for a fixed $z$,
equation (\ref{eq7.1}) becomes
\begin{eqnarray}
   \chi_s = \chi + \frac{1+z}{H(z)}(U - U_0),
\label{eq10}
\end{eqnarray}
where $U = |K|^{1/2} W$, $U_0 = |K|^{1/2} W_0$, for non-flat universe
and $U = W$, $U_0 = W_0$, for flat universe. With this expansion, the
density contrast $\delta^{\rm (s)}$ up to the first order is given by
\begin{eqnarray}
&&
   \delta^{\rm (s)}(\chi,\theta,\phi) =
   \frac{n^{\rm (s)}(\chi_s,\theta,\phi)}
      {\bar{\rho}\Phi(\chi_s,\theta,\phi)} - 1
\nonumber\\
&& \qquad\qquad =
   \delta(\chi,\theta,\phi) -
   \frac{\partial}{\partial\chi}
   \left( \frac{U}{aH} \right) - 
   \frac{A(\chi)}{aH}U +
   \left(
      \frac{A(\chi)}{aH} - \widetilde{q}_{\rm dec}
   \right) U_0,
\label{eq10.2}
\end{eqnarray}
where 
\begin{eqnarray}
   A(\chi) = \left\{
   \begin{array}{ll}
      \displaystyle
      \frac{\cosh\chi}{\sinh\chi}
      \left(
         2 + \frac{\partial\ln\Phi}{\partial\ln\sinh\chi}
      \right),
      & (K < 0) \\
      \displaystyle
      \frac{1}{\chi}
      \left(
         2 + \frac{\partial\ln\Phi}{\partial\ln\chi}
      \right),
      & (K = 0) \\
      \displaystyle
      \frac{\cos\chi}{\sin\chi}
      \left(
         2 + \frac{\partial\ln\Phi}{\partial\ln\sin\chi}
      \right),
      & (K > 0)
   \end{array}
   \right.
\label{eq10.5}
\end{eqnarray}
and $\widetilde{q}_{\rm dec}(z) = - d(a^{-1}H^{-1})/d\chi$, $\Omega =
8\pi G\rho/(3H^2)$ and $\lambda = \Lambda/(3H^2)$ are the normalized
time-dependent deceleration parameter, the time-dependent density
parameter and the dimensionless cosmological term, respectively. The
parameter $\widetilde{q}_{\rm dec}$ is equal to the usual
time-dependent deceleration parameter $\Omega/2 - \lambda$ for flat
universe, and $|K|^{-1/2}$ times the usual deceleration parameter for
non-flat universe. The light-cone effect of density contrast in
redshift space is represented by this equation (\ref{eq10.2}). It is
obvious that this equation is a generalization of corresponding
formula derived by Kaiser (1987) for $z \rightarrow 0$.

In the above equations, the light ray is assumed to be transmitted on
an unperturbed metric. In reality, the frequency of the light is
altered by the Sachs-Wolfe effect (Sachs \& Wolfe 1967), and the path
is bent by the gravitational lensing effect. The Sachs-Wolfe effect is
the contribution of the potential fluctuations to the estimate of the
redshift. The potential fluctuations are negligible except on scales
comparable to the Hubble distance, and in reality, observational
determination of the fluctuations on Hubble scales is not easy.
Therefore, the Sachs-Wolfe effect on the correlation function of
density field is not supposed to be important in practice.

The gravitational lensing changes the position on the sky of the
observable objects. The weak lensing (Kaiser 1992, 1998; Bernardeau,
Waerbeke, \& Mellier 1997) is relevant in our linear analysis. The
estimate of the number density is turn out to be affected by the local
convergence of the gravitational lensing. The local convergence is
given by the integration of the density fluctuations along the line of
sight (Bernardeau et al. 1997), and it is efficient for $z \simgt 1$.
Therefore, it is possible that the gravitational lensing affects the
correlation function for $z \simgt 1$. In which case, we should add a
correction term to the equation (\ref{eq10.2}). The quantitative
estimate of this effect is beyond the scope of this paper, and will be
investigated in a future paper.

\subsection{Redshift-space Distortion Operator}

The equations of motion relate the density contrast $\delta$ and the
velocity field $U$. Since we intend to include the curvature effect,
the Newtonian equations of motion are not appropriate. Instead, we
employ the perturbed Einstein equation,
\begin{eqnarray}
   \delta G^\mu_{\ \,\nu} = 8\pi G \delta T^\mu_{\ \,\nu},
\label{eq10.8}
\end{eqnarray}
on a background FLRW metric. In order to avoid the inclusion of
spurious gauge modes in the solution, Bardeen (1980) introduced the
gauge-invariant formalism for the above perturbed equation (see also
Kodama \& Sasaki 1984; Abbott \& Schaefer 1986; Hwang \& Vishniac
1990). In Appendix A, we review the gauge-invariant formalism for the
pressureless fluid, and derive complete equations to determine the
evolution of the density contrast and the velocity field. They are
given by equations (\ref{eqgi11a})--(\ref{eqgi11c}):
\begin{eqnarray}
&&
   \dot{\delta} + (\triangle + 3K) \psi = 0,
\label{eq11a}\\
&& \dot{\psi} + 2H \psi + \Phi = 0,
\label{eq11b}\\
&&
   (\triangle + 3K)\Phi = \frac32 H^2 \Omega \delta,
\label{eq11c}
\end{eqnarray}
where a dot denotes the differentiation with respect to the proper
time, $d/dt$, and $\triangle = \nabla^i \nabla_i$ is the Laplacian on
the 3-metric, $\gamma_{ij}$ of the FLRW metric, which is explicitly
given in Appendix B, equation (\ref{eqa1}). As shown in Appendix A,
the transverse part of the velocity field decays with time as $a^{-2}$
and can be neglected, thus the velocity is characterized by the
velocity potential $\psi$ so that the velocity is given by $v^i =
\nabla^i \psi$. The variables $\delta$, $\psi$, and $\Phi$ in the
above equations are actually gauge-invariant linear combinations
defined in Appendix A, and are guaranteed not to have spurious gauge
mode in the solution. These variables correspond to density contrast,
velocity potential, and gravitational potential inside the particle
horizon. In general, $\delta$ and $v$ correspond to the density
contrast and velocity in velocity-orthogonal isotropic gauge, $V=B$,
$H_T = 0$, in the notation of Appendix A. It is obvious that equations
(\ref{eq11a})--(\ref{eq11c}) correspond to the continuity, Euler, and
Poisson equations in Newtonian linear theory. The only difference is
the appearance of the curvature term in the Laplacian. The curvature
term would not be important in practice because the correlation
function on curvature scales is too small to be practically
detectable. However, we retain the curvature term for theoretical
consistency in the following.

Eliminating the variables $\psi$ and $\Phi$ from the equations, we
obtain the evolution equation for the Bardeen's gauge-invariant
density contrast $\delta$, given by
\begin{eqnarray}
   \ddot{\delta} + 2H \dot{\delta} -
   \frac32 H^2 \Omega \delta = 0.
\label{eq11.1}
\end{eqnarray}
This equation is equivalent to that in Newtonian theory for the
density contrast, and the solution of this equation is well-known
(Peebles 1980). The time dependence of the growing solution is given
by
\begin{eqnarray}
   D(t) \propto
   a\Omega \int_0^1
   \frac{dx}{(\Omega/x + \lambda x^2 + 1 - \Omega - \lambda)^{3/2}}.
\label{eq11.2}
\end{eqnarray}
In the following, we normalize this growing factor as $D(t_0) = 1$,
where $t_0$ is the present time. Thus, the growing solution of the
equation (\ref{eq11.1}) is given by
\begin{eqnarray}
   \delta(\chi,\theta,\phi,t) = 
   D(t) \delta_0(\chi,\theta,\phi),
\label{eq11.3}
\end{eqnarray}
where $\delta_0 \equiv \delta(t = t_0)$ is the mass density contrast
at the present time.

Equation (\ref{eq11a}) gives the solution of the velocity potential as
\begin{eqnarray}
   \psi =  - H D f (\triangle + 3K)^{-1} \delta_0,
\label{eq11.4}
\end{eqnarray}
where $f = (a/D)dD/da = \dot{D}/(HD)$ and $H =\dot{a}/a$ is the Hubble
parameter at time $t$. The inverse operator $(\triangle + 3K)^{-1}$ is
evaluated by spectral decomposition in the next section. It turns out
that the function $f$ depends only on $\Omega$ and $\lambda$, and is
approximately given by (Lahav et al. 1991)
\begin{eqnarray}
   f = \Omega^{0.6} + 
   \frac{\lambda}{70}\left( 1 + \frac{\Omega}{2} \right).
\label{eq11.5}
\end{eqnarray}
From equation (\ref{eq11.4}),
\begin{eqnarray}
   W = a n^i \nabla_i \psi =
   a H D f \frac{\partial}{\partial x} (\triangle + 3K)^{-1}
   \delta_0.
\label{eq12}
\end{eqnarray}
Since $\partial/\partial x = |K|^{1/2} \partial/\partial\chi$ for
non-flat universe and $\partial/\partial x = \partial/\partial\chi$
for flat universe, the above equation is equivalent to
\begin{eqnarray}
   U(\chi,\theta,\phi) = a H D f \times
   \left\{
   \begin{array}{ll}
      |K| \partial_\chi (\triangle + 3K)^{-1}
      \delta_0(\chi,\theta,\phi),
      & (K \ne 0) \\
      \partial_\chi \triangle^{-1}
      \delta_0(\chi,\theta,\phi),
      & (K = 0)
   \end{array}
   \right.
\label{eq12.5}
\end{eqnarray}
where $\partial_\chi = \partial/\partial\chi$.

Equations (\ref{eq10.2}) and (\ref{eq12.5}) indicate the linear
operator which transforms the density contrast at present in real
space to that in redshift space on a light-cone. That is, we can
define the redshift distortion operator:
\begin{eqnarray}
   \widehat{R} =
   1 + \beta \times
   \left\{
   \begin{array}{ll}
      |K| \left(\partial_\chi + \alpha\right)
      \partial_\chi (\triangle + 3K)^{-1},
      & (K \ne 0) \\
      \left(\partial_\chi + \alpha\right)
      \partial_\chi \triangle^{-1},
      & (K = 0)
   \end{array}
   \right.
\label{eq13}
\end{eqnarray}
where
\begin{eqnarray}
   \alpha(\chi) = \left\{
   \begin{array}{ll}
      \displaystyle
      \frac{\cosh\chi}{\sinh\chi}
      \left(
         2 +
         \frac{\partial\ln(Df\Phi)}{\partial\ln\sinh\chi}
      \right),
      & (K < 0) \\
      \displaystyle
      \frac{1}{\chi}
      \left(
         2 +
         \frac{\partial\ln(Df\Phi)}{\partial\ln\chi}
      \right),
      & (K = 0) \\
      \displaystyle
      \frac{\cos\chi}{\sin\chi}
      \left(
         2 +
         \frac{\partial\ln(Df\Phi)}{\partial\ln\sin\chi}
      \right).
      & (K > 0)
   \end{array}
   \right.
\label{eq14}
\end{eqnarray}
The time-dependent redshift distortion parameter is defined by
$\beta(z) = f(z)/b(z)$, where $b(z)$ is the time-dependent bias
parameter. In the following, we present the result in the case that
there is neither stochasticity nor scale-dependence in the biasing. It
is straightforward to include the scale-dependence and stochasticity,
although the expression becomes more tedious. In linear regime,
however, the stochasticity is shown to asymptotically vanish on large
scales and biasing is scale-independent beyond the scale of galaxy
formation, except some special cases (Matsubara 1999).

For the surveys of nearby universe, $z \ll 1$, the functions $D$ and
$f$ do not vary as rapidly as the selection function, $\Phi$, and the
factor $Df$ in the above equations can be omitted, as is done in the
literatures. However, if the selection function varies as slowly as
the factor $Df$, the latter factor cannot be neglected. With the
redshift distortion operator, we can re-express the density contrast
in redshift space as
\begin{eqnarray}
   \delta^{\rm (s)}(x,\theta,\phi) = 
   b D \widehat{R}\delta_0(x,\theta,\phi) + 
   \left(
      \frac{1 + z}{H(z)} A(\chi) - q_{\rm dec}(z)
   \right) U_0.
\label{eq15}
\end{eqnarray}
This expression depends on a peculiar motion of observer, $U_0$, which
somewhat complicate the analysis. Since we know the value $U_0$ by
measuring the dipole anisotropy of the CMB radiation (Kogut et al.
1993; Lineweaver et al. 1996), we can subtract the term which depends
on $U_0$ from the above expression. In the following we derive the
correlation function in redshift space from the observer in the CMB
frame and drop the $U_0$ term. In proper comparison of our result
below and the observation, one should be sure that the correlation
function is corrected by such kind of transformation to the CMB frame
(Hamilton 1998).

The correlation function in redshift space of two points, $\bfx_1$,
$\bfx_2$ is thus given by
\begin{eqnarray}
   \xi^{\rm (s)}(\bfx_1,\bfx_2) =
   b_1 b_2 D_1 D_2 \widehat{R}_1 \widehat{R}_2 \xi(\chi).
\label{eq16}
\end{eqnarray}
where $b_1 = b(z_1)$, $b_2 = b(z_2)$, $D_1 = D(z_1)$, $D_2 = D(z_2)$,
and $z_1$, $z_2$ are redshifts of the two points, and $\xi(\chi)$ is
the mass correlation function in real space at present time. The
redshift distortion operators $\widehat{R}_1$ and $\widehat{R}_2$
operate the density contrast at points $\bfx_1$ and $\bfx_2$,
respectively, and $\chi$ is a comoving separation of the two points.
If we do not omit the local velocity term, $U_0$, the square of the
second term in equation (\ref{eq15}) is added to the equation
(\ref{eq16}). In the following, we obtain an explicit expression for
equation (\ref{eq16}).

\section{THE CORRELATION FUNCTION IN REDSHIFT SPACE IN
FRIEDMAN-LEMA\^ITRE UNIVERSES} 
\label{sec3}
\setcounter{equation}{0}

In this section, we derive the explicit expression for the correlation
function in redshift space. We separately consider an open universe, a
flat universe and a closed universe in the following subsections.

\subsection{An Open Universe}

In an open universe ($K<0$), the correlation function in real space is
given by equation (\ref{eqa18ab}):
\begin{eqnarray}
&&
   \xi(\chi) = 
   \int\frac{\nu^2d\nu}{2\pi^2}
   X_0(\nu,\chi) S(\nu),
\label{eq18}
\end{eqnarray}
where
\begin{eqnarray}
   X_0 = \frac{\sin\nu\chi}{\nu\sinh\chi},
\label{eq18.5}
\end{eqnarray}
and $S(\nu)$ is the power spectrum of the gauge-invariant density
contrast (see Appendix B for detail)\footnote{Note that the power
spectrum near the horizon scale depends on the gauge choice. The power
spectrum $S(\nu)$ is defined by Bardeen's gauge-invariant density
contrast, i.e., the density contrast in a gauge with
velocity-orthogonal slicing, $V=B$.}. In this subsection, we omit the
superscript $(-)$ of $X_l$, which distinguish the difference of
functional forms according to the sign of curvature in Appendix B.
Equation (\ref{eq16}) implies that we only need to calculate
$\widehat{R}_1 \widehat{R}_2 X_0(\nu,\chi)$ to obtain the formula for
correlation function in redshift space. Since the Laplacian is the
invariant operator under transformation of the coordinate system, the
inverse operation $(\triangle_1 + 3K)^{-1}$, $(\triangle_2 + 3K)^{-1}$
to $X_0$ is simply given by $(-q^2 + 3K)^{-1} X_0 = - |K|^{-1}(\nu^2 +
4)^{-1} X_0$, where $q$ is defined by equation (\ref{eqa3a}), because
$X_0$ is the fundamental eigenfunction of the Laplacian [Appendix B,
equation (\ref{eqa2})].

\begin{figure}
\epsscale{0.6} \plotone{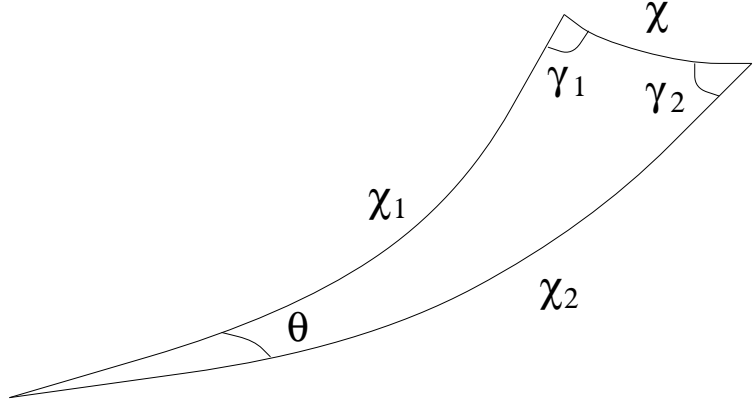}\\ \figcaption[fig1.ps]{Geometry of
the relative position of the observer and the two points. The space is
not necessarily Euclidean.
\label{fig1}}
\end{figure}

The calculationally nontrivial part of the redshift-space distortion
operator is the spatial derivatives along the lines of sight,
$\partial_\chi$, which is not invariant operation, unlike the
Laplacian. We illustrate in the rest of this subsection how the
calculation can be accomplished. The comoving separation $\chi$
between $\bfx_1$ and $\bfx_2$ is related to the comoving distances
$\chi_1$ and $\chi_2$ of these two points from the observer and the
angle, $\theta$ between the lines of sight of these points with
respect to the observer (Figure \ref{fig1}), through the standard
relation:
\begin{eqnarray}
   \cosh\chi = 
   \cosh\chi_1\cosh\chi_2 - \sinh\chi_1\sinh\chi_2\cos\theta.
\label{eq19}
\end{eqnarray}
\placefigure{fig1}
The derivatives of this equation yield
\begin{eqnarray}
&&
   \frac{\partial\chi}{\partial\chi_1} =
   \frac{1}{\sinh\chi}
   \left(
      \sinh\chi_1\cosh\chi_2 - \cosh\chi_1\sinh\chi_2\cos\theta
   \right) =
   \cos\gamma_1,
\label{eq20a}\\
&&
   \frac{\partial\chi}{\partial\chi_2} =
   \frac{1}{\sinh\chi}
   \left(
      \cosh\chi_1\sinh\chi_2 - \sinh\chi_1\cosh\chi_2\cos\theta
   \right) =
   \cos\gamma_2,
\label{eq20b}
\end{eqnarray}
where $\gamma_1$ is an angle between the geodesics of $\chi_1$ and
$\chi$, and $\gamma_2$ is an angle between the geodesics of $\chi_2$
and $\chi$ (Figure \ref{fig1}). The following equations are useful for
our purpose:
\begin{eqnarray}
&&
   \frac{\partial}{\partial\chi_1}
   \left( \sinh\chi \cos\gamma_1 \right) =
   \cosh\chi,
\label{eq20.1}\\
&&
   \frac{\partial}{\partial\chi_2}
   \left( \sinh\chi \cos\gamma_1 \right) =
   - \cosh\chi \cos\tth,
\label{eq20.2}
\end{eqnarray}
where 
\begin{eqnarray}
   \cos\tth = 
   \frac{\sin\gamma_1\sin\gamma_2}{\cosh\chi} -
   \cos\gamma_1\cos\gamma_2 =
   \frac{\cosh\chi_1\cosh\chi_2\cos\theta -
      \sinh\chi_1\sinh\chi_2}{\cosh\chi_1\cosh\chi_2 -
      \sinh\chi_1\sinh\chi_2\cos\theta}.
\label{eq20.4}
\end{eqnarray}
One can prove $|\cos\tth| \leq 1$ with this definition and $\theta
\rightarrow \tth$ for the scale much less than the curvature scale,
$\chi \ll |K|^{-2}$. We also use the derivatives of the radial part of
the harmonic function $X_l(\nu,\chi)$, given by equation
(\ref{eqa5a}), and the recursion relation, given by equation
(\ref{eqa6a}). The explicit form of the radial part of the harmonic
function $X_l$ is presented in Appendix B, equations
(\ref{eqa7a})--(\ref{eqa11a}).

\begin{figure}
\epsscale{1.0} \plotone{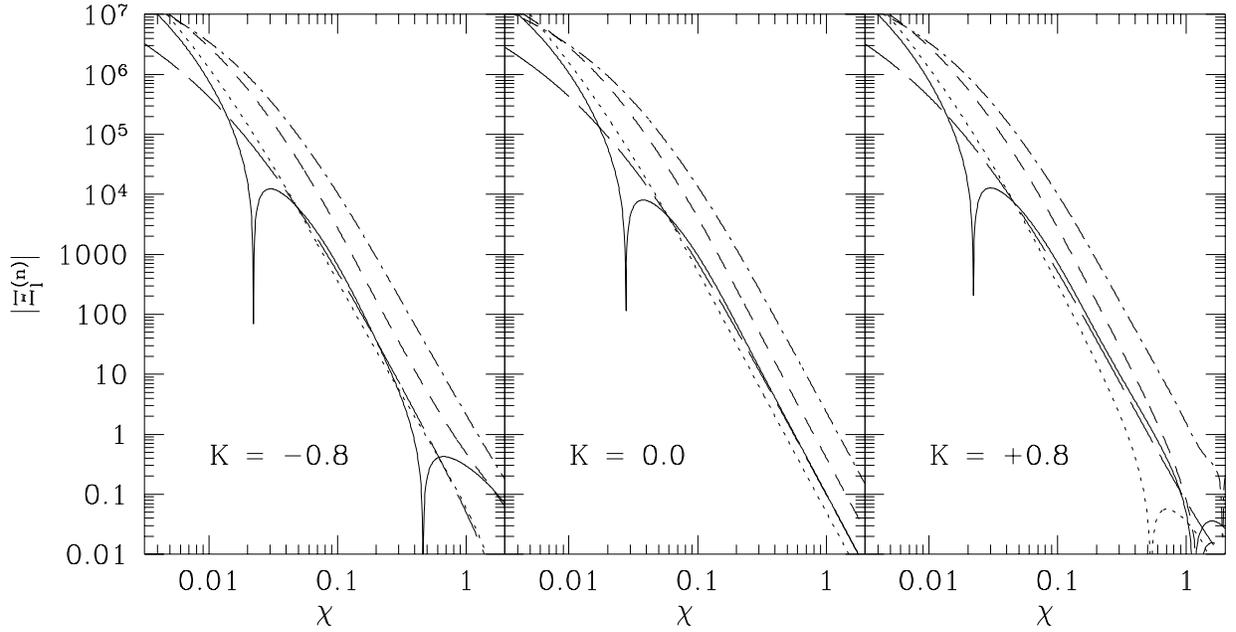}\\ \figcaption[fig2.eps]{Some examples
of the function $\Xi_l^{(n)}(\chi)$. The CDM-type power spectrum with
$\Gamma = 0.2 (\himpc)^{-1}$ and ``scale invariant'' (see text)
primordial spectrum is assumed and the normalization is arbitrary. The
curvature is $K = -0.8$, $0$, $+0.8$ from the left panel to the right
panel. Solid lines: $\Xi_0^{(0)}$, dotted lines: $\Xi_0^{(1)}$, dashed
lines: $\Xi_2^{(1)}$, long-dashed lines: $\Xi_2^{(2)}$, dash-dotted
lines: $\Xi_4^{(2)}$.
\label{fig2}}
\end{figure}

After straightforward but somewhat tedious algebra, using equations
(\ref{eq19})--(\ref{eq20.4}), (\ref{eqa5a}) and (\ref{eqa6a}), one
obtains the spatial derivatives along the lines of sight as
\begin{eqnarray}
&&
   \frac{\partial X_0}{\partial\chi_1} = 
   \cos\gamma_1 X_1,
\label{eq21}\\
&&
   \frac{\partial^2 X_0}{\partial\chi_1^{\,2}} = 
   X_0 -\frac13(\nu^2 + 4) X_0 +
   \left(\cos^2\gamma_1 - \frac13\right) X_2,
\label{eq22}\\
&&
   \frac{\partial^2 X_0}{\partial\chi_1\partial\chi_2} = 
   -\cos\tth X_0 + \frac13\cos\tth (\nu^2 + 4) X_0 +
   \left(
      \cos\gamma_1\cos\gamma_2 + \frac13 \cos\tth
   \right)
   X_2,
\label{eq23}\\
&&
   \frac{\partial^3 X_0}{\partial\chi_1^{\,2}\partial\chi_2} = 
   \cos\gamma_2 X_1 +
   \frac15 
   \left(
      2\cos\gamma_1\cos\tth - \cos\gamma_2 
   \right)
   (\nu^2 + 4) X_1
\nonumber\\
&&\qquad\qquad\quad +\,
   \frac15 
   \left(
      2\cos\gamma_1 \cos\tth + 
      5\cos^2\gamma_1\cos\gamma_2 - \cos\gamma_2
   \right)
   X_3,
\label{eq24}\\
&&
   \frac{\partial^4 X_0}{\partial\chi_1^{\,2}\partial\chi_2^{\,2}} = 
   X_0 -
   \frac{2}{15}
   \left(
      4 + 3\cos^2\tth
   \right)
   (\nu^2 + 4) X_0 +
   \frac{1}{15}
   \left(
      1 + 2\cos^2\tth
   \right)
   (\nu^2 + 4)^2 X_0
\nonumber\\
&&\qquad\qquad\quad -\,
   \frac{1}{21}
   \left[
      4 - 6\cos^2\tth -
      27\left(\cos^2\gamma_1 + \cos^2\gamma_2\right) -
      60\cos\gamma_1\cos\gamma_2\cos\tth
   \right]
   X_2
\nonumber\\
&&\qquad\qquad\quad +\,
   \frac{1}{21}
   \left[
      2 + 4\cos^2\tth -
      3\left(\cos^2\gamma_1 + \cos^2\gamma_2\right) +
      12\cos\gamma_1\cos\gamma_2\cos\tth
   \right]
   (\nu^2 + 4) X_2
\nonumber\\
&&\qquad\qquad\quad +\,
   \frac{1}{35}
   \left[
      1 + 2\cos^2\tth -
      5\left(\cos^2\gamma_1 + \cos^2\gamma_2\right) +
      20\cos\gamma_1\cos\gamma_2\cos\tth
   \right.
\nonumber\\
&&\qquad\qquad\qquad\qquad +\,
   \left.
      35\cos^2\gamma_1\cos^2\gamma_2
   \right]
   X_4.
\label{eq25}
\end{eqnarray}
Thus, $\widehat{R}_1 \widehat{R}_2 X_0(\nu,\chi)$ is expanded as
follows:
\begin{eqnarray}
   \widehat{R}_1 \widehat{R}_2 X_0(\nu,\chi) =
   \sum_{n,l} 
   c^{(n)}_l(\chi_1,\chi_2,\theta)
   \frac{(-1)^n X_l(\nu,\chi)}{\sinh^{2n -l}\chi(\nu^2 + 4)^n},
\label{eq27}
\end{eqnarray}
where $(n,l) = (0,0)$, $(1,0)$, $(1,1)$, $(1,2)$, $(2,0)$, $(2,1)$,
$(2,2)$, $(2,3)$, $(2,4)$, and coefficients $c^{(n)}_l$ are given by
\begin{eqnarray}
&&
   c^{(0)}_0 = 
   1 + \frac13(\beta_1 + \beta_2) +
   \frac{1}{15}\beta_1\beta_2
   \left(1 + 2\cos^2\tth\right),
\label{eq28}\\
&&
   c^{(1)}_0 = 
   \left[\beta_1 + \beta_2 +
      \frac{2}{15}\beta_1\beta_2\left(4 + 3\cos\tth\right)
   \right] \sinh^2\chi -
   \frac13 \beta_1 \beta_2 
   \widetilde{\alpha}_1\widetilde{\alpha}_2 \cos\tth,
\label{eq29}\\
&&
   c^{(1)}_1 = 
   \beta_1\widetilde{\alpha}_1\cos\gamma_1 +
   \beta_2\widetilde{\alpha}_2\cos\gamma_2
\nonumber\\
&&\qquad\quad + \,
   \frac15 \beta_1 \beta_2
   \left[
      \widetilde{\alpha}_1
      \left(\cos\gamma_1 - 2\cos\gamma_2\cos\tth\right) +
      \widetilde{\alpha}_2
      \left(\cos\gamma_2 - 2\cos\gamma_1\cos\tth\right)
   \right],
\label{eq30}\\
&&
   c^{(1)}_2 = 
   \beta_1 \left(\cos^2\gamma_1 - \frac13 \right) +
   \beta_2 \left(\cos^2\gamma_2 - \frac13 \right)
\nonumber\\
&&\qquad\quad  - \,
   \frac17 \beta_1 \beta_2
   \left[
      \frac23 + \frac43\cos^2\tth -
      \left(\cos^2\gamma_1 + \cos^2\gamma_2\right) +
      4\cos\gamma_1\cos\gamma_2\cos\tth
   \right],
\label{eq31}\\
&&
   c^{(2)}_0 = \beta_1 \beta_2
   \left( \sinh^2\chi -
      \widetilde{\alpha}_1 \widetilde{\alpha}_2
   \right) \sinh^2\chi,
\label{eq31.5}\\
&&
   c^{(2)}_1 = 
   \beta_1 \beta_2
   \left(
      \widetilde{\alpha}_1 \cos\gamma_1 +
      \widetilde{\alpha}_2 \cos\gamma_2
   \right) \sinh^2\chi,
\label{eq32}\\
&&
   c^{(2)}_2 = 
   \frac27 \beta_1\beta_2
   \left[
      \cos^2\tth - \frac23 + 
      \frac92 \left(\cos^2\gamma_1 + \cos^2\gamma_2\right) +
      10 \cos\gamma_1\cos\gamma_2\cos\tth
   \right] \sinh^2\chi
\nonumber\\
&&\qquad\quad  + \,
   \beta_1\beta_2\widetilde{\alpha}_1\widetilde{\alpha}_2
   \left(\cos\gamma_1\cos\gamma_2 + \frac13 \cos\tth\right),
\label{eq33}\\
&&
   c^{(2)}_3 = 
   \frac15 \beta_1 \beta_2
   \left[
      \widetilde{\alpha}_1
      \left(
         5\cos\gamma_1\cos^2\gamma_2 - \cos\gamma_1 +
         2\cos\gamma_2\cos\tth
      \right)
   \right.
\nonumber\\
&&\qquad\qquad\qquad
   \left. + \,
      \widetilde{\alpha}_2
      \left(
         5\cos\gamma_2\cos^2\gamma_1 - \cos\gamma_2 +
         2\cos\gamma_1\cos\tth
      \right)
   \right],
\label{eq34}\\
&&
   c^{(2)}_4 = 
   \frac17 \beta_1\beta_2
   \left[
      \frac15 + \frac25\cos^2\tth -
      \left(\cos^2\gamma_1 + \cos^2\gamma_2\right) +
      4\cos\gamma_1\cos\gamma_2\cos\tth
   \right.
\nonumber\\
&&\qquad\qquad\qquad  + \,
   \left.
      7\cos^2\gamma_1\cos^2\gamma_2
   \right],
\label{eq35}
\end{eqnarray}
where
\begin{eqnarray}
&&
   \widetilde{\alpha}_1(\chi_1,\chi) =
   \alpha(\chi_1) \sinh\chi =
   \sinh\chi \frac{\cosh\chi_1}{\sinh\chi_1}
   \left(
      2 +
      \frac{\partial\ln(D_1 f_1 \Phi_1)}{\partial\ln\sinh\chi_1}
   \right),
\label{eq35.5a}\\
&&
   \widetilde{\alpha}_2(\chi_2,\chi) =
   \alpha(\chi_2) \sinh\chi =
   \sinh\chi \frac{\cosh\chi_2}{\sinh\chi_2}
   \left(
      2 +
      \frac{\partial\ln(D_2 f_2 \Phi_2)}{\partial\ln\sinh\chi_2}
   \right),
\label{eq35.5b}
\end{eqnarray}
and $\beta_1 = \beta(z_1)$, $\beta_2 = \beta(z_2)$, $f_1 = f(z_1)$,
$f_2 = f(z_2)$, $\Phi_1 = \Phi(\chi_1)$, $\Phi_2 = \Phi(\chi_2)$.
Since these coefficients do not depend on $\nu$, the correlation
function in redshift space of equation (\ref{eq16}) finally reduces to
\begin{eqnarray}
   \xi(\bfx_1,\bfx_2) = 
   b_1 b_2 D_1 D_2 \sum_{n,l} 
   c^{(n)}_l(\chi_1,\chi_2,\theta)
   \Xi^{(n)}_l(\chi),
\label{eq36}
\end{eqnarray}
where
\begin{eqnarray}
   \Xi^{(n)}_l(\chi) =
    \frac{(-1)^n}{\sinh^{2n-l}\chi}
   \int \frac{\nu^2d\nu}{2\pi^2}
   \frac{X_l(\nu,\chi)}{(\nu^2 + 4)^n} 
   S(\nu).
\label{eq37}
\end{eqnarray}
In the left panel of Figure \ref{fig2}, some examples of the function
$\Xi^{(n)}_l(\chi)$ for an open model are plotted. The CDM-type
transfer function (Bardeen et al. 1986) is adopted and the primordial
power spectrum on large scales is assumed to be ``scale invariant'',
which corresponds to constant fluctuations in the gravitational
potential per logarithmic interval in wavenumber (Lyth \& Stewart
1990; Ratra \& Peebles 1994; White \& Bunn 1995):
\begin{eqnarray}
   S(\nu) \propto \frac{(\nu^2 + 4)^2}{\nu(\nu^2 + 1)}
      T^2\left(|K|^{1/2}\nu/\Gamma\right),
\label{eq38.01}
\end{eqnarray}
where
\begin{eqnarray}
   T(p) =
   \frac{\ln (1 + 2.34 p)}{2.34 p}
   \left[
      1 + 3.89 p + (16.1 p)^2 + (5.46 p)^3 + (6.71 p)^4
   \right]^{-1/4},
   \label{eq38.02}
\end{eqnarray}
and the shape parameter is set $\Gamma/(\himpc = 0.2)$, which
corresponds to $\Gamma = 6 \times 10^2 (cH_0^{\,-1})^{-1}$ in our unit
system. On scales smaller than Horizon scale, $\nu \gg 1$, the power
spectrum (\ref{eq38.01}) reduces to usual CDM-type power spectrum with
Harrison-Zel'dovich primordial spectrum.

\placefigure{fig2}

The set of equations (\ref{eq28})--(\ref{eq37}) is our final formula
for the correlation function in redshift space which takes into
account the light-cone, and spatial-curvature effect {\em without}
distant-observer approximation.

\subsection{A Flat Universe}

In a flat universe ($K=0$), the correlation function is real space is
given by equation (\ref{eqa18ab}):
\begin{eqnarray}
&&
   \xi(\chi) = 
   \int\frac{\nu^2d\nu}{2\pi^2}
   X_0(\nu,\chi) S(\nu),
\label{eq38.11}
\end{eqnarray}
where
\begin{eqnarray}
   X_0 = \frac{\sin\nu\chi}{\nu\chi}.
\label{eq38.12}
\end{eqnarray}
We can repeat the similar calculation of the previous subsection for a
flat universe ($K=0$). The corresponding formula can also be used if
we only consider a nearby universe where the separation $\chi$ and the
distances $\chi_1$, $\chi_2$ are much smaller than the curvature
scale, $|K|^{-1/2}$, which roughly corresponds to the horizon scale.

The calculation for a flat universe is performed similarly as in the
previous subsection, or alternatively, we can also obtain the formula
in a flat limit from the formula for an open universe, taking the
limit $\chi_1, \chi_2, \chi \rightarrow 0$, $\nu \rightarrow \infty$,
with $\nu\chi$ fixed. Anyway, the equation (\ref{eq19}) reduces to the
one in Euclidean geometry:
\begin{eqnarray}
   \chi^2 = \chi_1^{\,2} + \chi_2^{\,2} - 2 \chi_1 \chi_2 \cos\theta. 
\label{eq38.1}
\end{eqnarray}
The meaning of $\gamma_1$ and $\gamma_2$ is the same, and they are
explicitly given by
\begin{eqnarray}
   \cos\gamma_1 = \frac{\chi_1 - \chi_2\cos\theta}{\chi},
\qquad
   \cos\gamma_2 = \frac{\chi_2 - \chi_1\cos\theta}{\chi}.
\label{eq38.5}
\end{eqnarray}
The variable $\tth$ in flat universe, defined by equation
(\ref{eq20.4}) for the case of an open universe, reduces to $\theta$,
the angle between lines of sight of $\bfx_1$ and $\bfx_2$ with respect
to the observer.

Thus, the correlation function in redshift space is given by equation
(\ref{eq36}) where the coefficients $c_l^{(n)}$ in a flat universe are
\begin{eqnarray}
&&
   c^{(0)}_0 = 
   1 + \frac13(\beta_1 + \beta_2) +
   \frac{1}{15}\beta_1\beta_2
   \left(1 + 2\cos^2\theta\right),
\label{eq39}\\
&&
   c^{(1)}_0 = 
   -\frac13 \beta_1 \beta_2 
   \widetilde{\alpha}_1\widetilde{\alpha}_2 \cos\theta,
\label{eq40}\\
&&
   c^{(1)}_1 = 
   \beta_1\widetilde{\alpha}_1 \cos\gamma_1 +
   \beta_2\widetilde{\alpha}_2 \cos\gamma_2
\nonumber\\
&&\qquad\quad + \,
   \frac15 \beta_1 \beta_2
   \left[
      \widetilde{\alpha}_1
      \left(\cos\gamma_1 - 2\cos\gamma_2\cos\theta\right) +
      \widetilde{\alpha}_2
      \left(\cos\gamma_2 - 2\cos\gamma_1\cos\theta\right)
   \right],
\label{eq41}\\
&&
   c^{(1)}_2 = 
   \beta_1 \left(\cos^2\gamma_1 - \frac13 \right) +
   \beta_2 \left(\cos^2\gamma_2 - \frac13 \right)
\nonumber\\
&&\qquad\quad  - \,
   \frac17 \beta_1 \beta_2
   \left[
      \frac23 + \frac43\cos^2\theta -
      \left(\cos^2\gamma_1 + \cos^2\gamma_2\right) +
      4\cos\gamma_1\cos\gamma_2\cos\theta
   \right],
\label{eq42}\\
&&
   c^{(2)}_0 = 0,
\label{eq42.5}\\
&&
   c^{(2)}_1 = 0,
\label{eq43}\\
&&
   c^{(2)}_2 = 
   \beta_1\beta_2\widetilde{\alpha}_1\widetilde{\alpha}_2
   \left(\cos\gamma_1\cos\gamma_2 + \frac13 \cos\theta\right),
\label{eq44}\\
&&
   c^{(2)}_3 = 
   \frac15 \beta_1 \beta_2
   \left[
      \widetilde{\alpha}_1
      \left(
         5\cos\gamma_1\cos^2\gamma_2 - \cos\gamma_1 +
         2\cos\gamma_2\cos\theta
      \right)
   \right.
\nonumber\\
&&\qquad\qquad\qquad
   \left. + \,
      \widetilde{\alpha}_2
      \left(
         5\cos\gamma_2\cos^2\gamma_1 - \cos\gamma_2 +
         2\cos\gamma_1\cos\theta
      \right)
   \right],
\label{eq45}\\
&&
   c^{(2)}_4 = 
   \frac17 \beta_1\beta_2
   \left[
      \frac15 + \frac25\cos^2\theta -
      \left(\cos^2\gamma_1 + \cos^2\gamma_2\right) +
      4\cos\gamma_1\cos\gamma_2\cos\theta
   \right.
\nonumber\\
&&\qquad\qquad\qquad  + \,
   \left.
      7\cos^2\gamma_1\cos^2\gamma_2
   \right],
\label{eq46}
\end{eqnarray}
where
\begin{eqnarray}
&&
   \widetilde{\alpha}_1(\chi_1,\chi) =
   \alpha(\chi_1) \chi =
   \frac{\chi}{\chi_1}
   \left(
      2 +
      \frac{\partial\ln(D_1 f_1 \Phi_1)}{\partial\ln\chi_1}
   \right),
\label{eq46.5a}\\
&&
   \widetilde{\alpha}_2(\chi_2,\chi) =
   \alpha(\chi_2) \chi =
   \frac{\chi}{\chi_2}
   \left(
      2 +
      \frac{\partial\ln(D_2 f_2 \Phi_2)}{\partial\ln\chi_2}
   \right),
\label{eq46.5b}
\end{eqnarray}
The corresponding equation of (\ref{eq37}) is
\begin{eqnarray}
&&
   \Xi^{(n)}_l(\chi) =
   \frac{(-1)^n}{\chi^{2n-l}}
   \int \frac{\nu^2d\nu}{2\pi}
   \frac{X_l(\nu,\chi)}{\nu^{2n}} 
   S(\nu) =
   \frac{(-1)^{n+l}}{x^{2n-l}}
   \int \frac{k^2dk}{2\pi^2}
   \frac{j_l(kx)}{k^{2n-l}} P(k).
\label{eq47}
\end{eqnarray}
In the middle panel of Figure \ref{fig2}, some examples of the
function $\Xi^{(n)}_l(\chi)$ for a flat model are plotted for scale
invariant primordial power spectrum with CDM-type transfer function:
\begin{eqnarray}
   S(\nu) \propto \nu T^2(\nu/\Gamma),
\label{eq47.01}
\end{eqnarray}
where we again set the shape parameter $\Gamma/(\himpc) = 0.2$.

Alternatively, the coefficients $c_l^{(n)}$ can also be represented by
using variables $\gamma \equiv \gamma_2 + \theta$ and $\thh =
\theta/2$. These variables are introduced in Szalay, Matsubara \&
Landy (1998) in a calculation of wide-angle effects of nearby universe
($z \ll 1$), except that their original definition of $\gamma$ is $\pi
- \gamma$, instead. The meaning of $\gamma$ is the angle between
$\bfx_2 - \bfx_1$ and a symmetry axis that halves the angle $\theta$.

With these new variables, equations (\ref{eq39})--(\ref{eq46}) are
transformed as
\begin{eqnarray}
&&
   c^{(0)}_0 = 
   1 + \frac13(\beta_1 + \beta_2) +
   \frac{1}{5}\beta_1\beta_2 -
   \frac{8}{15}\beta_1\beta_2\cos^2\thh\sin^2\thh,
\label{eq48}\\
&&
   c^{(1)}_0 = 
   - \frac13 \beta_1 \beta_2 
   \widetilde{\alpha}_1\widetilde{\alpha}_2 \cos2\thh,
\label{eq49}\\
&&
   c^{(1)}_1 = 
   \left[
      (\beta_1 \widetilde{\alpha}_1 + \beta_2 \widetilde{\alpha}_2) +
      \frac15 \beta_1\beta_2
      (\widetilde{\alpha}_1 + \widetilde{\alpha_2})
      \left(
         3 - 4\cos^2\thh
      \right)
   \right]
   \sin\thh\sin\gamma
\nonumber\\
&&\qquad\quad  -\,
   \left[
      \left(
         \beta_1 \widetilde{\alpha}_1 - \beta_2 \widetilde{\alpha}_2
      \right) +
      \frac15 \beta_1\beta_2
      \left(
         \widetilde{\alpha}_1 - \widetilde{\alpha}_2
      \right)
      \left(
         3 - 4\sin^2\thh
      \right)
   \right]
   \cos\thh\cos\gamma,
\label{eq50}\\
&&
   c^{(1)}_2 = 
   \left[
      \frac23 (\beta_1 + \beta_2) + \frac47\beta_1\beta_2
   \right]
   \cos2\thh P_2(\cos\gamma)
\nonumber\\
&& \qquad\quad + \,
   \frac13
   \left[
      (\beta_1 + \beta_2) - \frac27\beta_1\beta_2 +
      \frac87\beta_1\beta_2\sin^2\thh
   \right]
   \sin^2\thh
\nonumber\\
&& \qquad\quad - \,
   2(\beta_1 - \beta_2)\cos\thh\sin\thh\cos\gamma\sin\gamma,
\label{eq51}\\
&&
   c^{(2)}_0 = 0,
\label{eq51.5}\\
&&
   c^{(2)}_1 = 0,
\label{eq52}\\
&&
   c^{(2)}_2 = 
   \frac13\beta_1\beta_2
   \widetilde{\alpha}_1\widetilde{\alpha}_2
   \left(\sin^2\thh - 2P_2(\cos\gamma)\right),
\label{eq53}\\
&&
   c^{(2)}_3 = 
   - \frac15 \beta_1\beta_2
   \left\{
      \left(
         \widetilde{\alpha}_1 +
         \widetilde{\alpha}_2
      \right)
      \sin\thh
      \left[
         \cos^2\thh P_1(\sin\gamma) - 2P_3(\sin\gamma)
      \right]
   \right.
\nonumber\\
&&\qquad\qquad\qquad\quad - \,
   \left.
      \left(
         \widetilde{\alpha}_1 -
         \widetilde{\alpha}_2
      \right)
      \cos\thh
      \left[
         \sin^2\thh P_1(\cos\gamma) - 2P_3(\cos\gamma)
      \right]
   \right\},
\label{eq54}\\
&&
   c^{(2)}_4 = 
   \frac17\beta_1\beta_2
   \left[
      \frac85 P_4(\cos\gamma) -
      \frac43 \sin^2\thh P_2(\cos\gamma) -
      \frac{1}{15}(4 - 9\sin^2\thh)\sin^2\thh
   \right],
\label{eq55}
\end{eqnarray}
where $P_l(x)$ is the Legendre function.

\subsection{A Closed Universe}

In a closed universe ($K>0$), the correlation function in real space
is given by the equation (\ref{eqa18c}):
\begin{eqnarray}
&&
   \xi(\chi) = 
   \sum_{\nu=3}^\infty \frac{\nu^2}{2\pi^2}
   X_0(\nu,\chi) S(\nu),
\label{eq55.5}
\end{eqnarray}
where
\begin{eqnarray}
   X_0 = \frac{\sin\nu\chi}{\nu\sin\chi}.
\label{eq55.6}
\end{eqnarray}
Repeating the similar calculation as in an open universe, or formally
putting $\chi \rightarrow i\chi$, $\chi_1 \rightarrow i\chi_1$,
$\chi_2 \rightarrow i\chi_2$ and $\nu \rightarrow -i\nu$ in the
calculation for an open universe, we obtain the formula for a closed
universe. In this subsection, we just summarize the difference of the
formula from the case of an open universe.

The corresponding equation (\ref{eq19}) in a closed universe is
\begin{eqnarray}
   \cos\chi = 
   \cos\chi_1\cos\chi_2 + \sin\chi_1\sin\chi_2\cos\theta,
\label{eq56}
\end{eqnarray}
The meaning of $\gamma_1$ and $\gamma_2$ is unchanged and they are
given by
\begin{eqnarray}
&&
   \cos\gamma_1 =
   \frac{1}{\sin\chi}
   \left(
      \sin\chi_1\cos\chi_2 - \cos\chi_1\sin\chi_2\cos\theta
   \right),
\label{eq57a}\\
&&
   \cos\gamma_2 =
   \frac{1}{\sin\chi}
   \left(
      \cos\chi_1\sin\chi_2 - \sin\chi_1\cos\chi_2\cos\theta
   \right)
\label{eq57b}
\end{eqnarray}
The variable $\tth$ in a closed universe is defined by
\begin{eqnarray}
   \cos\tth = 
   \frac{\sin\gamma_1\sin\gamma_2}{\cos\chi} -
   \cos\gamma_1\cos\gamma_2 =
   \frac{\cos\chi_1\cos\chi_2\cos\theta +
      \sin\chi_1\sin\chi_2}{\cos\chi_1\cos\chi_2 +
      \sin\chi_1\sin\chi_2\cos\theta}.
\label{eq58}
\end{eqnarray}
Then the form of the coefficients $c_l^{(n)}$ of equation
(\ref{eq28})--(\ref{eq35}) is almost the same, and we do not repeat
the formula here. The only difference is that $\sinh\chi$ should be
replaced by $\sin\chi$, and the definition of $\widetilde{\alpha}$ is
changed as
\begin{eqnarray}
&&
   \widetilde{\alpha}_1(\chi_1,\chi) =
   \alpha(\chi_1) \sin\chi =
   \sin\chi \frac{\cos\chi_1}{\sin\chi_1}
   \left(
      2 +
      \frac{\partial\ln(D_1 f_1 \Phi_1)}{\partial\ln\sin\chi_1}
   \right),
\label{eq58.5a}\\
&&
   \widetilde{\alpha}_2(\chi_2,\chi) =
   \alpha(\chi_2) \sin\chi =
   \sin\chi \frac{\cos\chi_2}{\sin\chi_2}
   \left(
      2 +
      \frac{\partial\ln(D_2 f_2 \Phi_2)}{\partial\ln\sin\chi_2}
   \right).
\label{eq58.5b}
\end{eqnarray}
The formula for a closed universe is also given by the equation
(\ref{eq36}) where the definition of $\Xi^{(n)}_l$ is replaced by
\begin{eqnarray}
   \Xi^{(n)}_l(\chi) =
   \frac{1}{\sin^{2n-l}\chi}
   \sum_{\nu=3}^\infty \frac{\nu^2}{2\pi^2}
   \frac{X_l(\nu,\chi)}{(\nu^2 - 4)^n} 
   S(\nu).
\label{eq59}
\end{eqnarray}
In the right panel of Figure \ref{fig2}, some examples of the function
$\Xi^{(n)}_l(\chi)$ are plotted for a closed model. As in the open
case, the primordial power spectrum on large scales is assumed to be
``scale invariant'', which corresponds to constant fluctuations in the
gravitational potential per logarithmic interval in wavenumber (White
\& Scott 1996):
\begin{eqnarray}
   S(\nu) \propto \frac{(\nu^2 - 4)^2}{\nu(\nu^2 - 1)}
      T^2\left(|K|^{1/2}\nu/\Gamma\right).
\label{eq60.01}
\end{eqnarray}
where we again set the shape parameter as $\Gamma/(\himpc) = 0.2$

\subsection{Recovery of the Known Formulas of the Redshift Distortions
of the Correlation Function}

It is an easy exercise to derive the previously known formulas of the
redshift distortions of the correlation function from our general
formula. We illustrate how our formula reduces to the known formulas
by taking appropriate limits. Here we consider two approximations,
$\chi \ll \chi_1, \chi_2$ and/or $z \ll 1$. The first approximation
corresponds to the distant observer approximation. In this
approximation, the distance between two points is much smaller than
the distances of these points from the observer. The second
approximation corresponds to only considering the nearby universe. The
redshift distortion formulas known so far are restricted in these
cases.

First of all, consider a limit $\chi \ll \chi_1, \chi_2$, with
$\gamma_1$, $\gamma_2$ fixed. We also set $\beta_1 = \beta_2 = \beta$,
and the distance between the two points $\chi$ is much smaller than
the curvature scale. Then irrespective to the spatial curvature,
$\theta, \tth \rightarrow 0$, and $\gamma_2 \rightarrow \pi - \gamma_1
\equiv \gamma$.  Both the coefficients (\ref{eq28})--(\ref{eq35}), and
(\ref{eq39})--(\ref{eq46}) reduce to
\begin{eqnarray}
&&
   c^{(0)}_0 = 
   1 + \frac23\beta +
   \frac{1}{5}\beta^2
\label{eq61}\\
&&
   c^{(1)}_2 = 
   \left(
      \frac43 \beta + \frac47 \beta^2
   \right) P_2(\cos\gamma),
\label{eq62}\\
&&
   c^{(2)}_4 = 
   \frac{8}{35} \beta^2 P_4(\cos\gamma),
\label{eq63}
\end{eqnarray}
and all the other coefficients are zero. These coefficients are
equivalent to the result that Hamilton (1992) derived in $z
\rightarrow 0$ limit with distant observer approximation, which is a
direct Fourier transform of Kaiser's original form in Fourier space
(Kaiser 1987). For finite $z$, the equivalent result is obtained by
Matsubara \& Suto (1996) for correlation function and Ballinger,
Peacock \& Heavens (1996) for power spectrum [see also Nakamura,
Matsubara \& Suto (1998) de Laix \& Starkman (1998), Nair (1999)]. All
these previous studies are based on the distant observer
approximation, and our general formula correctly has the limit of
these cases.

The wide-angle effects on the redshift distortion of the correlation
function without distant observer approximation is already derived for
nearby universe, $z \ll 0$ by Szalay, Matsubara \& Landy (1998), and
Bharadwaj (1999) also derived the equivalent result with another
parameterization. It is easy to see that our general results have the
correct limit of Szalay et al. (1998). In the limit $z\rightarrow 0$
the geometry reduces to be flat case, and, in fact, equations
(\ref{eq48})--(\ref{eq55}) with $\beta_1 = \beta_2 \equiv \beta$ in
that limit is completely equivalent to the result of Szalay et al.
(1998) after correcting their typographical errors\footnote{The factor
$4/15$ in the last term of their equation (15) should be replaced by
$8/15$ and the left hand side of their equation (20) should be
replaced by $1/3 \cdot \alpha_1 \alpha_2 \beta^2 \cos 2\theta$.},
noting the minor difference of the definition that $\gamma$ and
$\Xi^{(n)}_l$ here correspond to $\pi - \gamma$ and $(-1)^{n+l}
x^{-2n+l} \xi^{(2n-l)}_l$ in Szalay et al. (1998), respectively.

\section{THE REDSHIFT DISTORTIONS AND THE SPATIAL CURVATURE OF
THE UNIVERSE}
\label{sec4}
\setcounter{equation}{0}

It is well known that the redshift distortions of the correlation
function for nearby universe are good probes of the density parameter
modulo bias factor at present time, $\beta_0 = \Omega_0^{\,0.6}/b_0$.
The parameter $\beta_0$ is called as the redshift distortion parameter
of nearby universe. Redshift distortions of nearby universe do not
depend on spatial curvature almost at all, because the redshift
distortions of nearby universe are purely caused by the peculiar
velocity field, which is almost independent on the spatial curvature.
However, the redshift distortions at high redshifts, say $z \simgt 1$
in a quasar catalog, definitely depend on the spatial curvature of the
universe and is called as cosmological redshift distortion. Matsubara
\& Suto (1996) explicitly show the dependence of the cosmological
distortion of the correlation function on cosmological parameters,
employing the distant observer approximation. To probe more properly
the spatial curvature with cosmological distortions, it is not
desirable to rely on the distant observer approximation.

In this section, we numerically calculate the formula obtained in the
previous section for several models, concentrating on geometrical
effects. We plot the correlation function in directly observable
velocity space. In the following, the shape parameter for the CDM-type
transfer function is fixed to $\Gamma = 6 \times 10^2$ in our unit
system, which corresponds to $\Gamma/(\himpc) = 0.2$, in spite of the
fact that CDM model predict $\Gamma/(\himpc) = \Omega_0 h$. We fix the
spectrum simply because we are interested in the pure distortion
effects on the difference among the models, while the difference of
the shape of the underlying power spectrum is less interested in here.
The primordial spectrum is assumed to be ``scale invariant'', which
corresponds to constant fluctuations in the gravitational potential
per logarithmic interval in wavenumber, equations (\ref{eq38.01}),
(\ref{eq47.01}) and (\ref{eq60.01}). For simplicity, we assume no
bias, $b = 1$ and also assume $\alpha = 0$. The latter assumption
corresponds to the case $Df\Phi(\chi) \propto (\sinh\chi)^{-2}$,
$\chi^{-2}$, and $(\sin\chi)^{-2}$, for open, flat, and closed models,
respectively. Although this form of selection function is not
physically motivated, it is not so unrealistic for merely illustrative
purpose. In actual application, the selection function is individually
determined for each redshift survey.

\begin{figure}
\epsscale{1.0} \plotone{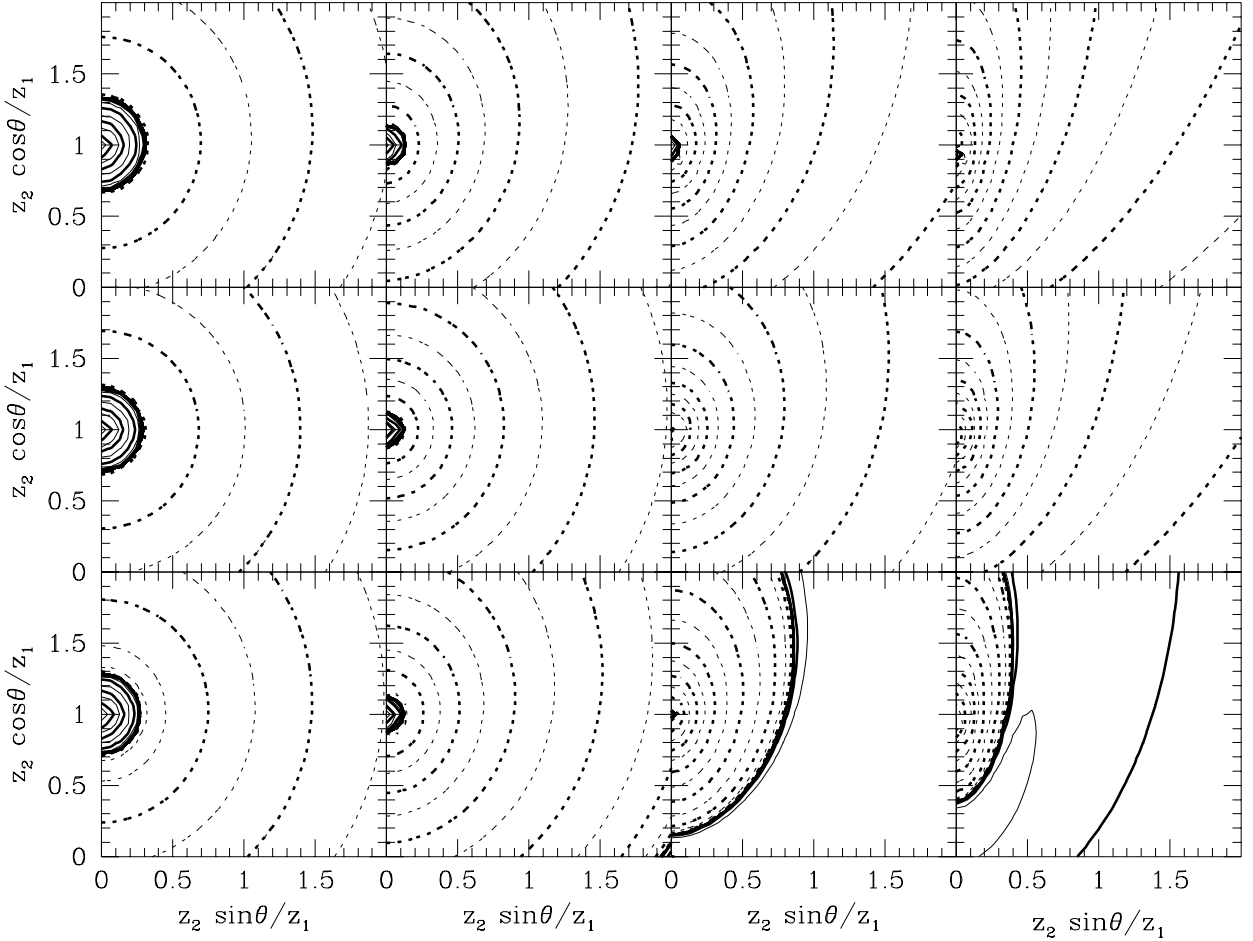}\\ \figcaption[fig3.eps]{Contour
plots of the correlation function in real space with purely
geometrical distortions, without velocity distortions. The observer is
sitting at the origin, and the first point at redshift $z_1$ is
sitting at the center on the y-axis. The contour map shows the value
of the correlation function depending on the position of the second
point at redshift $z_2$ and angle $\theta$. The solid lines indicate
positive correlation and the dotted lines indicate the negative
correlation. The normalization is arbitrary and the contour spacings
are $\Delta\log_{10}|\xi| = 0.5$. The top panels are for the STD model
($\Omega_0=1$, $\lambda_0=0$), the middle for the FLAT model
($\Omega_0 = 0.2$, $\lambda_0 = 0.8$), and the bottom for the OPEN
model ($\Omega_0 = 0.2$, $\lambda_0 = 0$). From left to right panels,
the redshifts of the first point $z_1$ are 0.1, 0.3, 1, 3,
respectively.
\label{fig3}}
\end{figure}

\begin{figure}
\epsscale{1.0} \plotone{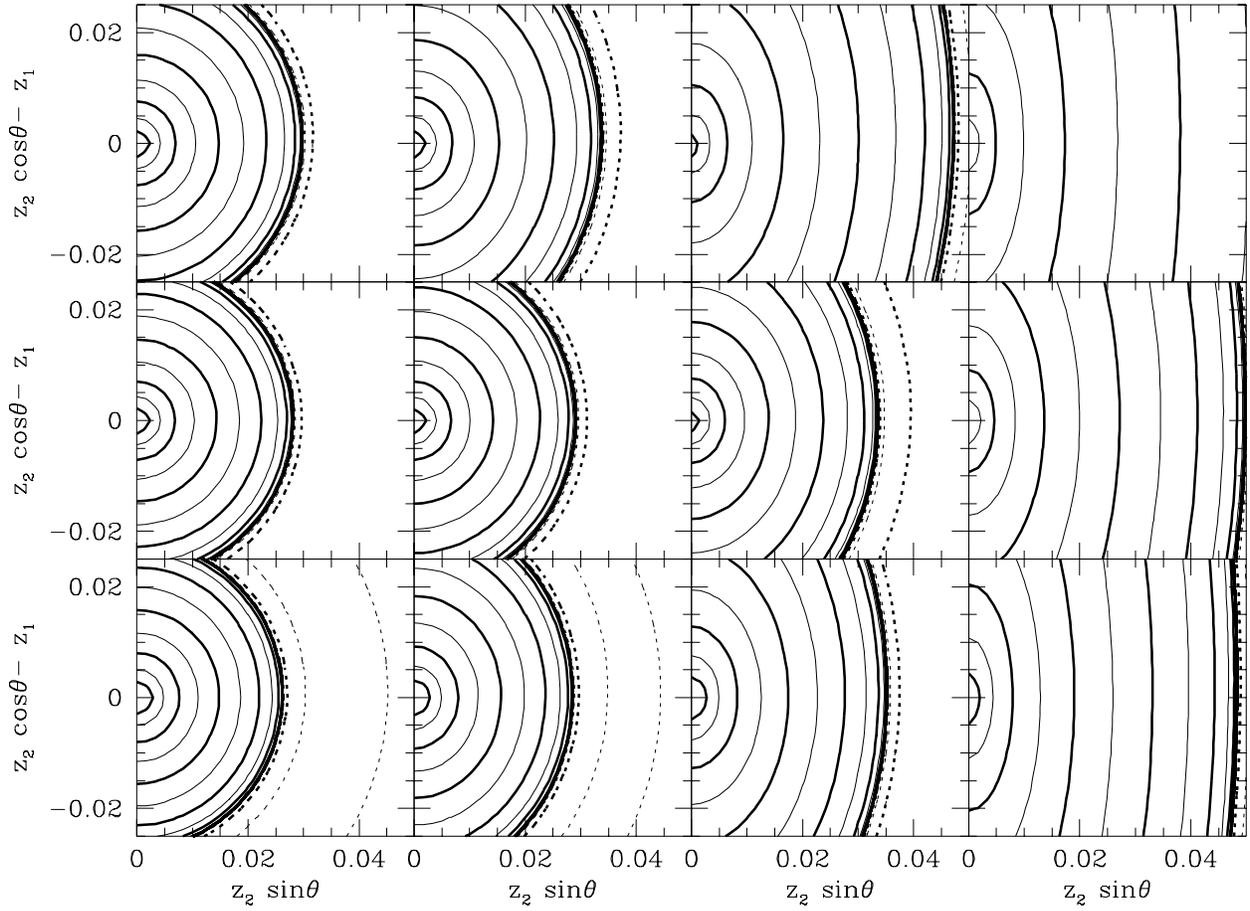}\\ \figcaption[fig4.eps]{Same as Figure
\ref{fig3}, but the scale of $z$ around the first point is
fixed. The observer is located at $(0, -z_1)$, which is outside of the
plots.
\label{fig4}}
\end{figure}

\begin{figure}
\epsscale{1.0} \plotone{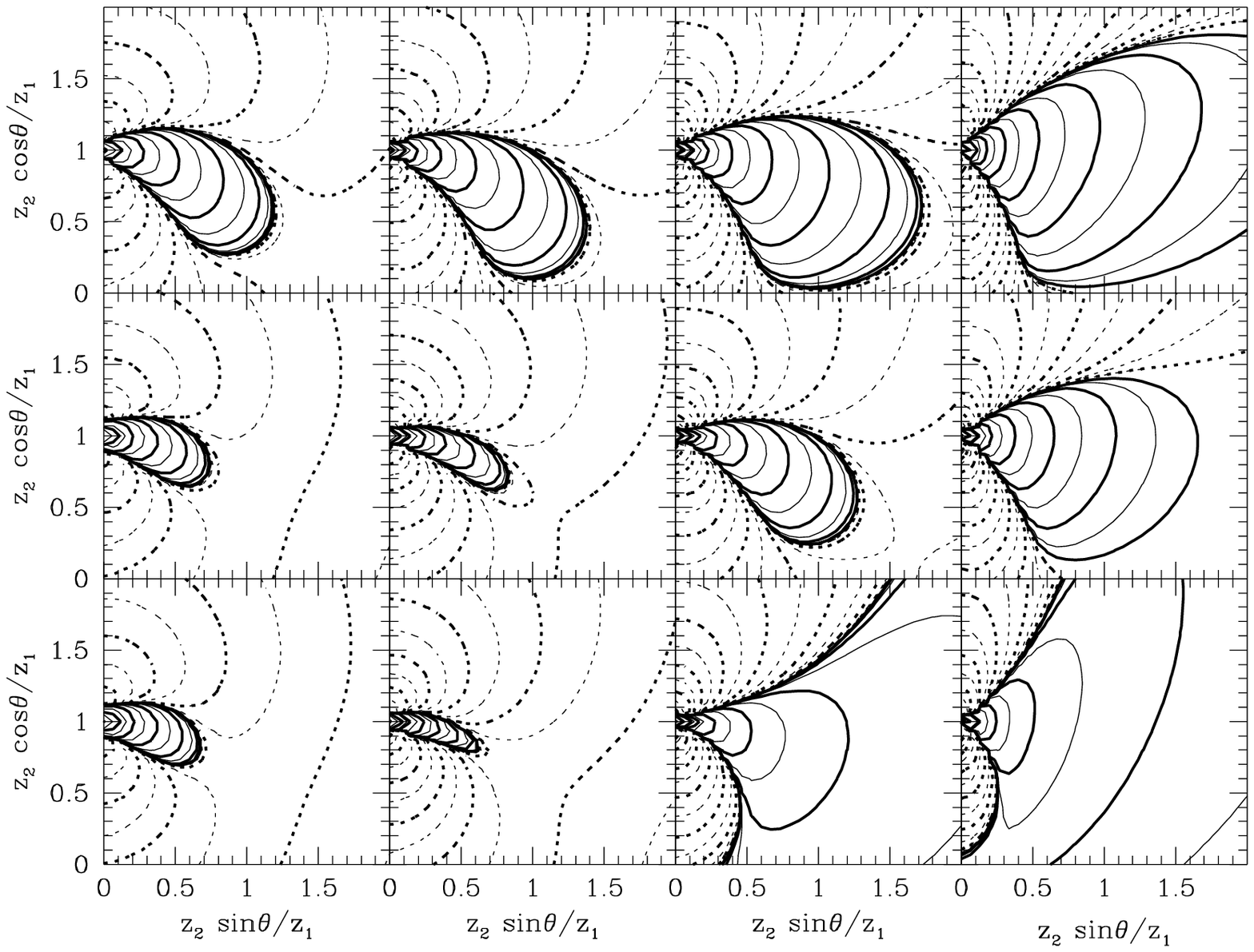}\\ \figcaption[fig5.eps]{Contour plots
of the correlation function in redshift space. The meaning of the
panels are the same as in Figure \ref{fig3}, but the velocity
distortions are included.
\label{fig5}}
\end{figure}

\begin{figure}
\epsscale{1.0} \plotone{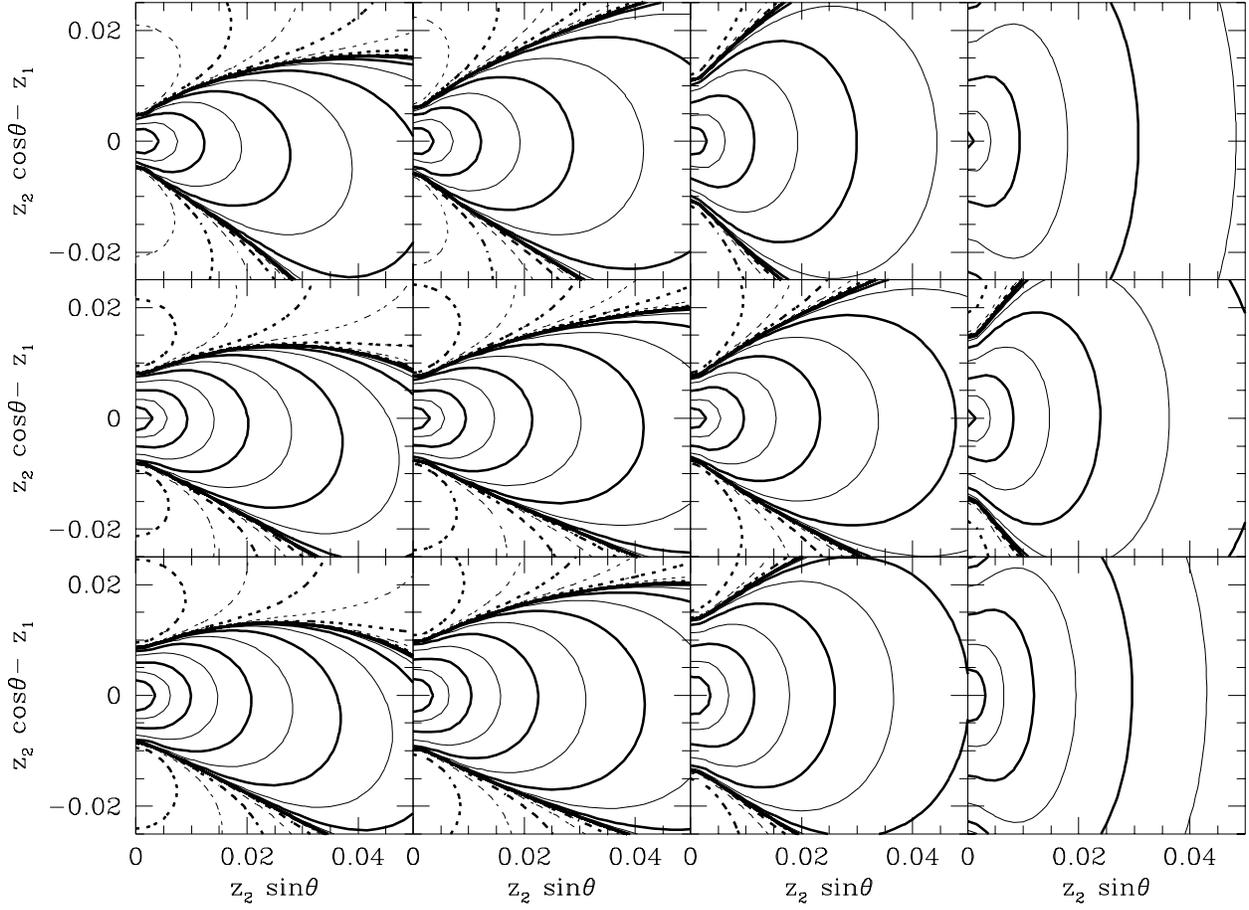}\\ \figcaption[fig6.eps]{Contour plots
of the correlation function in redshift space with fixed scale of $z$.
The meaning of the panels are the same as in Figure \ref{fig4}, but
the velocity distortions are included.
\label{fig6}}
\end{figure}

{}Figures \ref{fig3}--\ref{fig6} show the contour plots of the
correlation function. Each figure consists of 12 panels. In each
figure, from top to bottom, the cosmological models are $(\Omega_0,
\lambda_0) = (1, 0), (0.2, 0.8), (0.2, 0)$, respectively, which we
call STD, FLAT, and OPEN models. From left to right, the redshifts of
the first points are $z = 0.1, 0.3, 1.0, 3.0$, respectively.

\placefigure{fig3}
\placefigure{fig4}
\placefigure{fig5}
\placefigure{fig6}

In Figures \ref{fig3} and \ref{fig4}, the correlation function with
the purely geometrical distortions is plotted for an illustrative
purpose, assuming there are no peculiar velocities at all, by setting
$\beta = 0$ in our formula. The first point, $\bfx_1$ of equation
(\ref{eq36}), is at the center on the y-axis in each figures. The
contour plot show the value of correlation function depending on the
position of the second point $\bfx_2$. In Figure \ref{fig3}, the
observer is located at the origin $(0, 0)$ and the global distortions
are shown. In Figure \ref{fig4}, the scale of $z$ around the first
object is fixed, and the observer is located at $(0, -z_1)$, outside
the plots.

{}For lower redshifts, $z = 0.1, 0.3$, the geometrical distortion is
not significant. For higher redshifts, $z = 1.0, 3.0$, due to the
nonlinear relations of redshift and comoving distance, the contours
are elongated to the direction of line of sight in all three models.
The extent of the elongation for STD and OPEN models are similar, but
the elongation for FLAT model is smaller than other models. This is
because the acceleration nature due to the cosmological constant
squashes the z-space along the line of sight (Alcock \& Paczy\'nski
1979).

In Figures \ref{fig5} and \ref{fig6}, the correlation function in
redshift space is plotted, taking into account the velocity
distortions. As in Figures \ref{fig3} and \ref{fig4}, the observer is
located $(0, 0)$ and $(0, - z_1)$, respectively.

{}For lower redshifts, the redshift distortions are mainly from
peculiar velocity fields, which depends only on the density parameter
$\Omega_0$. Thus, STD model can be discriminated from other models by
correlation function of lower redshifts, but FLAT and OPEN models are
similar. For higher redshifts, FLAT and OPEN models become different
because of the squashing by the cosmological constant.

\begin{figure}
\epsscale{0.5} \plotone{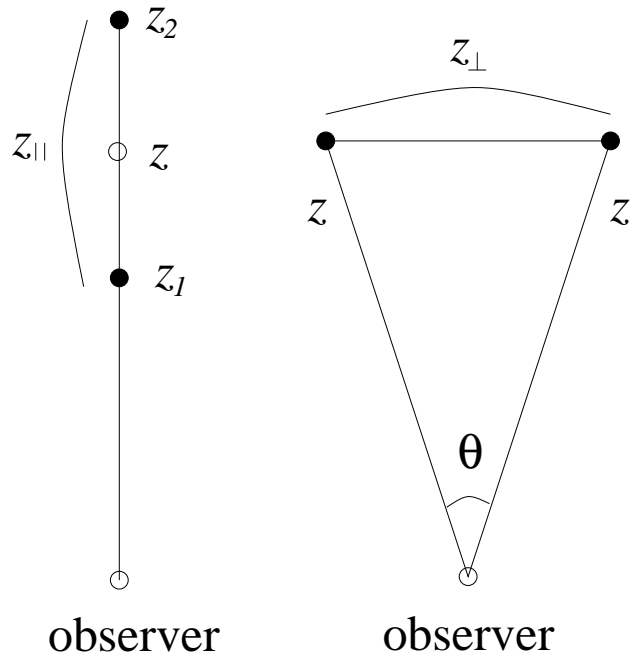}\\ \figcaption[fig7.eps]{Geometrical
meaning of the definition of the parallel redshift interval $z_\Vert$
and the perpendicular redshift interval $z_\bot$.
\label{fig7}}
\end{figure}

To illustrate the difference among profiles of correlation function in
redshift space for different models and different redshifts, we define
the parallel correlation function $\xi_\Vert(z_\Vert;z)$ and
perpendicular correlation function $\xi_\bot(z_\bot;z)$ at given $z$.
In terms of the correlation function in redshift space $\xi^{\rm
(s)}(z_1, z_2, \theta)$ they are defined by
\begin{eqnarray}
&&
   \xi_\Vert(z_\Vert; z) =
   \xi^{\rm (s)}
      \left(z - \frac{z_\Vert}{2}, z + \frac{z_\Vert}{2}, 0\right),
\label{eq64}\\
&&
   \xi_\bot(z_\bot; z) =
   \xi^{\rm (s)}
      \left(z, z, 
         {\rm Arccos}\left[1 - \frac{z_\bot^{\,2}}{2 z^2}\right]
      \right).
\label{eq65}
\end{eqnarray}
The geometrical meaning of the definition of the parallel redshift
interval $z_\Vert$ and the perpendicular redshift interval $z_\bot$
are illustrated in Figure \ref{fig7}. As seen from the Figure
\ref{fig5} or \ref{fig6}, these sections of correlation function are
supposed to be maximally distorted in opposite direction, and that is
the reason why we introduce them for illustrations. These functions
are the generalization of the similar functions introduced by
Matsubara \& Suto (1996) in the case of distant observer
approximation.

\placefigure{fig7}

\begin{figure}
\epsscale{1.0} \plotone{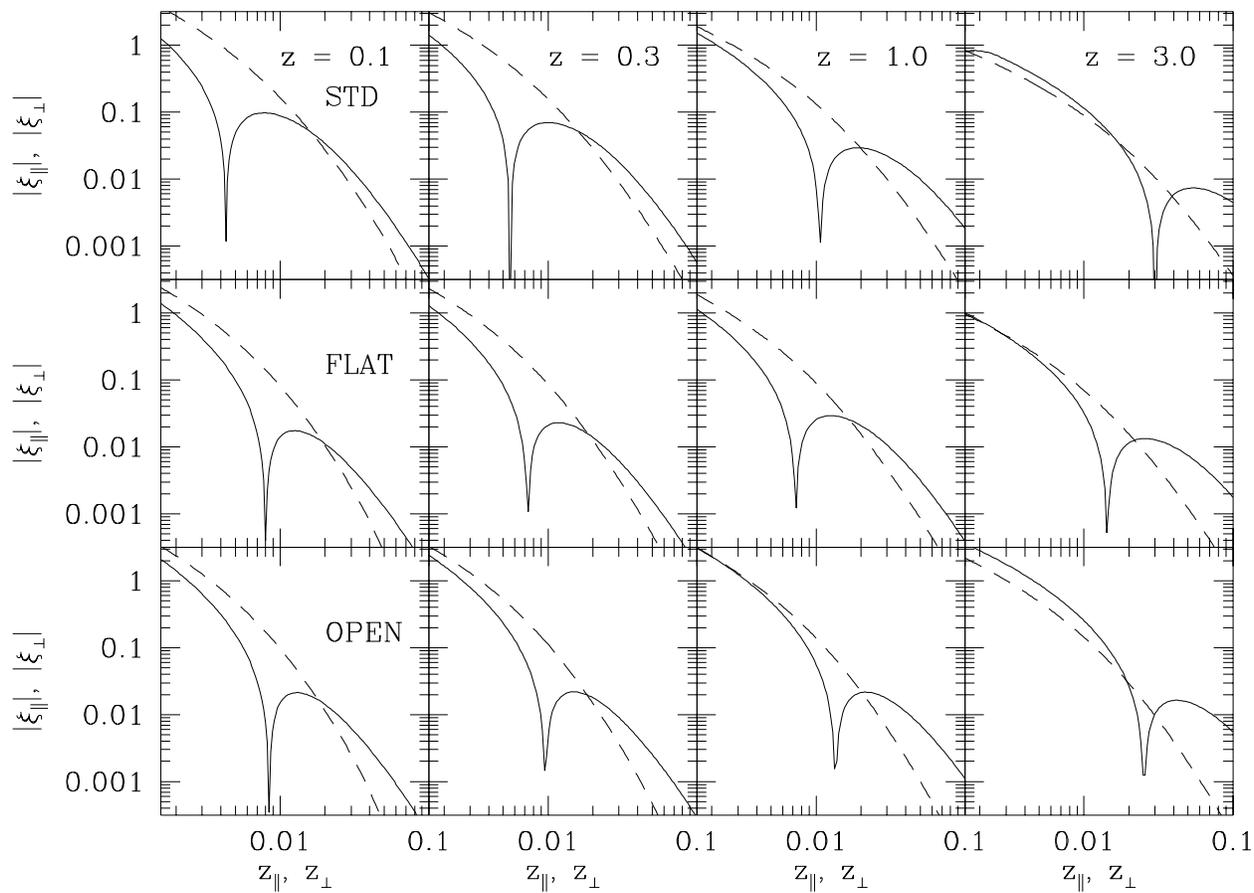}\\ \figcaption[fig8.ps]{The parallel
correlation function, $\xi_\Vert$ (solid lines), and the perpendicular
correlation function, $\xi_\bot$ (dashed lines), in redshift space.
The redshifts of the first point are 0.1, 0.3, 1, 3 from left to right
panels. The cosmological models are STD, FLAT, and OPEN from top to
bottom panels.
\label{fig8}}
\end{figure}

In Figure \ref{fig8}, those parallel and perpendicular correlation
functions are plotted for STD, FLAT, and OPEN models. The correlation
function in real space is normalized as $\sigma_8 = 1$. One can notice
that while the perpendicular correlation functions are not
significantly different among three models, the parallel correlation
functions are quite different. For lower redshifts, profiles of the
parallel correlation function for the FLAT and OPEN models are similar
as usual. For higher redshifts, the relative amplitude of $\xi_\Vert$
compared to $\xi_\bot$ in the OPEN model is higher than that in the
FLAT model. In addition to that, the zero-crossing point of
$\xi_\Vert$ in the OPEN model is larger than that in the FLAT model.
Those tendencies are both explained by the cosmological-constant
squashing, because the squashing shifts the profile toward small
scales, or left. Even if the determination of the zero-crossing point
of $\xi_\Vert$ is observationally difficult, the former effect on the
relative amplitude is a promising one to discriminate the spatial
curvature by the observation of redshift distortions.

\placefigure{fig8}

\begin{figure}
\epsscale{1.0} \plotone{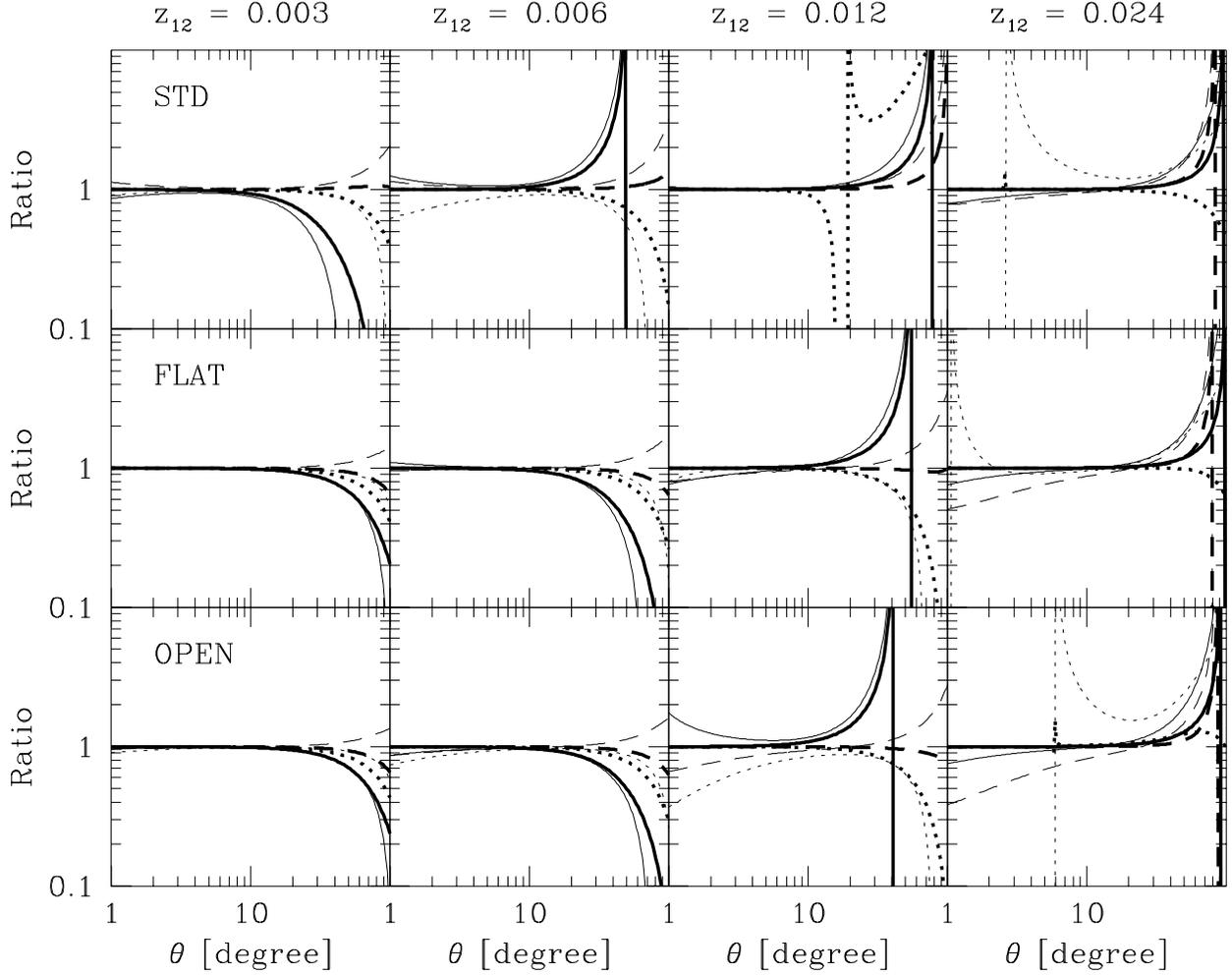}\\ \figcaption[fig9.ps]{The ratio of
the correlation functions by distant observer approximation to that of
our result. The inclination angle $\gamma_z$ (see text) is fixed to
$\gamma_z = 10^\circ$ (solid lines), $45^\circ$ (dotted lines), and
$80^\circ$ (dashed lines). The angle $\theta$ between the lines of
sight is varied. The redshift separation $z_{12}$ in velocity space is
fixed to 0.003, 0.006, 0.012, 0.024 from left to right panels. Thin
lines show the approximation by Hamilton (1992) for a nearby universe.
Thick lines show the approximation by Matsubara \& Suto (1996). Top,
middle, bottom panels show each different cosmological models, STD,
FLAT, and OPEN, respectively.
\label{fig9}}
\end{figure}

The distant observer approximation has been widely used in the
analyses of galaxy redshift surveys. With our general formula, we can
figure out when this approximation is valid and when it is not. The
distant observer approximation of the two-point correlation function
is derived by Hamilton (1992) for a nearby universe, $z \ll 1$.
Matsubara \& Suto (1996) generalize his formula to arbitrary
redshifts, in which the distant observer approximation is still
adopted. In Figure \ref{fig9}, we plot the ratio of the value of those
two previous formulas and that of our formula. We choose the geometry
of the two points as follows: we fix the separation $z_{12} \equiv
(z_1^{\,2} + z_2^{\,2} - 2 z_1 z_2 \cos\theta)^{1/2}$ of the two
points in velocity space. The angle $\gamma_ z$ between the symmetric
line which halves the lines of sight and the line between the two
points is also fixed. The meaning of the angle $\gamma_z$ is the
inclination of $z_{12}$ relative to the line of sight in velocity
space. Explicitly, $\gamma_z$ is given by $\tan\gamma_z = (z_1 +
z_2)|z_1 - z_2|^{-1}\tan(\theta/2)$. The angle $\theta$ between the
lines of sight is varied in the figure. We plot the cases, $z_{12} =
0.003$, $0.006$, $0.012$, $0.024$, and $\gamma_z = 10^\circ$,
$45^\circ$, $80^\circ$. There is some irregular behavior that
corresponds to the zero crossings of the correlation function.

\placefigure{fig9}

The Hamilton's formula (thin lines) is valid when the separation
$z_{12}$ is not so large. The reason why the Hamilton's formula
deviate even in small angles is that when the angle is small enough,
the redshift becomes large, and the evolutionary and geometrical
effects are not negligible. In fact, the formula of Matsubara \& Suto
(thick lines) is perfectly identical to our general formula when the
angle is small.

The validity region of the distant observer approximation depends on
the inclination angle $\gamma_z$. However, one can conservatively
estimate the validity region as $\theta \simlt 10^\circ$. One can use
$\theta \sim 20^\circ$ for some cases, while the blind application of
the approximation for $\theta \sim 30^\circ$ can cause over 100\%
error.

\section{CONCLUSIONS}
\label{sec5}
\setcounter{equation}{0}

In this paper we have derived for the first time the unified formula
of the correlation function in redshift space in linear theory, which
simultaneously includes wide-angle effects and the cosmological
redshift distortions. The effects of the spatial curvature both on
geometry and on fluctuation spectrum are properly taken into account,
and our formula applies to an arbitrary Friedman-Lema\^{\i}tre
universe.

The distant observer approximation by Matsubara \& Suto (1996), which
is a generalization of the work by Hamilton (1992), can be used when
the angle $\theta$ between the lines of sight is less than $10^\circ$.
Beyond that range, our formula provides a unique one for the
correlation function in redshift space with geometrical distortions.

The correlation function in redshift space is uniquely determined if
the cosmological parameters $\Omega_0$, $\lambda_0$, the power
spectrum $P(k)$ and bias evolution $b(z)$ is specified. The Hubble
constant does not affect the correlation function, provided that we do
not adopt a specific model to the power spectrum and/or to the bias
evolution which may depend on Hubble constant. Our formula predicts
the correlation function for a fixed model of these variables and one
can test any model by directly comparing the correlation function of
the data and of the theoretical prediction in redshift space.

Now we comment on some caveats for our formula. First, when we try to
apply our formula to the scale $\simlt 20\himpc$, what apparently
lacks in our formula is the finger-of-God effect, or the nonlinear
smearing of the correlation along the line of sight. While it is still
difficult to analytically include this effect into our formula, one
can phenomenologically evaluate the effect by numerically smearing the
formula along the line of sight. Second, our formula does not include
the Sachs-Wolfe and gravitational lensing effects. The Sachs-Wolfe
effect could affect our formula only on scales comparable to Hubble
distance where the correlation function is too small to be practically
detectable, as discussed in section \ref{sec2.1}. However, it is
possible that the gravitational lensing effect affects the observable
correlation function for $z \simgt 1$. In which case, one should add
correction terms to our formula. Even so, our formula for $z \simlt 1$
would not be affected by those terms and still corresponds to the
observable quantity. Those correction terms by gravitational lensing
will be given in a future paper. Third, our formula assumes the
selection function uncorrelated to the density fluctuations. In
addition, the precise form of the selection function is practically
difficult to determine. If the luminocities and/or the surface
brightnesses are correlated to density fluctuations, it can mimick the
large-scale structure.

It could be the case that one can blindly seek the models of
cosmological quantities, $\Omega_0$, $\lambda_0$, $P(k)$, and $b(z)$.
Each quantity has different effects on the correlation function in
redshift space. Roughly speaking, the cosmological parameters
$\Omega_0$ and $\lambda_0$ mostly affect the distortions of the
contour of the correlation function, the power spectrum $P(k)$ mostly
affects the profile of the correlation function, and the bias
evolution mostly affects the z-dependence of the amplitude. Since all
the pairs of the objects in the redshift survey can be used, we can
expect that those quantities are determined with small errors by
performing proper likelihood analysis, such as the Karhunen-Lo\`eve
mode decomposition (Vogeley \& Szalay 1996; Matsubara, Szalay \& Landy
1999). The application of the formula to the actual data is a
straightforward task. We believe the formula presented in this paper
is one of the most fundamental theoretical tools in understanding the
data of the deep redshift surveys.

\acknowledgements

I wish to thank Alex Szalay, Naoshi Sugiyama and Yasushi Suto for
stimulating discussions. This work was supported by JSPS Postdoctoral
Fellowships for Research Abroad.

\newpage

\appendix

\section*{APPENDIX A\\ GAUGE-INVARIANT EVOLUTION EQUATIONS FOR THE
DENSITY CONTRAST AND THE VELOCITY FIELD}

\renewcommand{\theequation}{\mbox{\rm
{\Alph{section}\arabic{equation}}}}  
\setcounter{section}{1}
\setcounter{equation}{0}

In this appendix, we review the derivation of the equations of motion
for the density contrast and the velocity field in the general
relativistic context. We assume the matter is pressureless, perfect
fluid, and the background metric is given by homogeneous, isotropic
FLRW metric. For the perturbed Einstein equation,
\begin{eqnarray}
   \delta G^\mu_{\ \,\nu} = 8\pi G \delta T^\mu_{\ \,\nu},
\label{eqgi1}
\end{eqnarray}
Bardeen (1980) introduced the gauge-invariant formalism. Here, we
briefly review the formalism in the case of pressureless fluid, and
derive relevant equations for our purpose, i.e., equations to relate
the density contrast and the velocity field. In this appendix, we use
the conformal time $\tau$ defined by $d\tau = dt/a$, so that the
metric is given by
\begin{eqnarray}
   ds^2 = a^2(\tau) \left(-d\tau^2 + \gamma_{ij}dx^i dx^j \right),
\label{eqgi2.5}
\end{eqnarray}
in the coordinates $(\tau, x^i)$. A prime denotes differentiation with
respect to the conformal time.

The perturbations are classified into scalar, vector, and tensor
types. The scalar perturbations are expanded by the complete set of
scalar harmonics $Q(x^i)$ satisfying the Helmholtz equation
\begin{eqnarray}
 (\triangle + q^2)Q = 0,
\label{eqgi2}
\end{eqnarray}
where $\triangle = \nabla^i \nabla_i$ is the Laplacian and $\nabla_i$
is the three-dimensional covariant derivative with respect to the
metric $\gamma_{ij}$. The explicit form of each mode of this solution
is given in Appendix B. In the following, the indices to distinguish
different modes are omitted and each mode function is represented by
$Q$.

For each mode, the scalar perturbations in the metric (\ref{eqgi2.5})
are defined by
\begin{eqnarray}
&&
   h_{00} = -2 a^2 A(\tau) Q,
\label{eqgi3a}\\
&&
   h_{0i} = - a^2 B(\tau) Q_i,
\label{eqgi3b}\\
&&
   h_{ij} =  2 a^2
   \left[ H_L(\tau) Q \gamma_{ij} + H_T(\tau) Q_{ij} \right],
\label{eqgi3c}
\end{eqnarray}
where
\begin{eqnarray}
&&
   Q_i = - \frac{1}{q} \nabla_i Q,
\label{eqgi7a}\\
&&
   Q_{ij} = 
   \frac{1}{q^2} \nabla_i \nabla_j Q + \frac13 \gamma_{ij} Q.
\label{eqgi7b}
\end{eqnarray}
The three-velocity $w^i$ associated with four-velocity $u^\mu$ is
represented by
\begin{eqnarray}
   w^i = \frac{u^i}{u^0} = V(\tau) Q^i,
\label{eqgi4}
\end{eqnarray}
and to first order the normalization $u_\mu u^\mu = -1$ gives
\begin{eqnarray}
   u^0 = \frac{1 - A(\tau) Q}{a}.
\label{eqgi5}
\end{eqnarray}
The pressureless energy-momentum tensor is given by $T^\mu_{\ \,\nu} =
\rho u^\mu u_\nu$, thus, the scalar perturbations in the pressureless
energy-momentum tensor are
\begin{eqnarray}
&&
   \delta T^0_{\ \,0} = - \bar{\rho} \Delta(\tau) Q,
\label{eqgi6a}\\
&&
   \delta T^0_{\ \,j} = \bar{\rho} [V(\tau) - B(\tau)] Q_j,
\label{eqgi6b}\\
&&
   \delta T^i_{\ \,j} = 0.
\label{eqgi6c}
\end{eqnarray}
In terms of the gauge-dependent variables $A$, $B$, $H_L$, $H_T$, $V$,
and $\Delta$, Bardeen (1980) defined gauge invariant combinations:
\begin{eqnarray}
&&
   \Phi_A = A + \frac{1}{q} B' + 
   \frac{1}{q} \frac{a'}{a} B -
   \frac{1}{q^2}\left({H_T}'' + \frac{a'}{a} {H_T}' \right),
\label{eqgi8a}\\
&&
   \Phi_H = H_L + \frac13 H_T +
   \frac{1}{q} \frac{a'}{a} B - 
   \frac{1}{q^2} \frac{a'}{a} {H_T}',
\label{eqgi8b}\\
&&
   \delta = \Delta + \frac{3}{q} \frac{a'}{a} (V - B),
\label{eqgi8c}\\
&&
    V_s = V - \frac{1}{q} {H_T}'.
\label{eqgi8d}
\end{eqnarray}
The perturbation equation (\ref{eqgi1}) for these scalar
gauge-invariant variables are given by Bardeen (1980):
\begin{eqnarray}
&&
  (q^2 - 3K) \Phi_H = 
  4\pi G a^2 \bar{\rho} \delta,
\label{eqgi9a}\\
&&
   \Phi_A + \Phi_H = 0,
\label{eqgi9b}\\
&&
   {V_s}' + \frac{a'}{a} V_s = q \Phi_A,
\label{eqgi9c}\\
&&
   \delta' + (q^2 - 3K)\frac{1}{q} V_s = 0.
\label{eqgi9d}
\end{eqnarray}
We define new variables $\psi = - a^{-1} q^{-1} V_s$ and $\Phi =
a^{-2}\Phi_A$. After superposing the modes of the above equations, we
obtain
\begin{eqnarray}
&&
   \dot{\delta} + (\triangle + 3K) \psi = 0,
\label{eqgi11a}\\
&& \dot{\psi} + 2H \psi + \Phi = 0,
\label{eqgi11b}\\
&&
   (\triangle + 3K)\Phi = \frac32 H^2 \Omega \delta,
\label{eqgi11c}
\end{eqnarray}
where a dot denotes the differentiation with respect to the proper
time, $d/dt$, and $\triangle$ is the Laplacian on FLRW metric,
equation (\ref{eqa1}), This is the complete set of equations to treat
the scalar perturbations.

The velocity $w^i$ corresponds to $dx^i/d\tau = adx^i/dt = a v^i$, and
the equation (\ref{eqgi4}) suggests $w^i = a\nabla^i \psi$, so that
$v^i = \nabla^i \psi$, i.e., $\psi$ is the velocity potential and
represents the longitudinal part of the velocity $v^i$. The
transverse part of $v^i$ belongs to the vector perturbations.

The vector perturbations are expanded by the divergenceless vector
harmonics $Q^{(1)}_i$ satisfying
\begin{eqnarray}
&&
   (\triangle + q^2)Q^{(1)}_i = 0,
\label{eqgi12a}\\
&&
   \nabla^i Q^{(1)}_i = 0.
\label{eqgi12b}
\end{eqnarray}
For each mode, the vector perturbations are given by
\begin{eqnarray}
&&
   h_{00} = 0,
\label{eqgi13a}\\
&&
   h_{0i} = - a^2 B^{(1)}(\tau) Q^{(1)}_i,
\label{eqgi13b}\\
&&
   h_{ij} =  2 a^2 H^{(1)}_T(\tau) Q^{(1)}_{ij},
\label{eqgi13c}
\end{eqnarray}
where
\begin{eqnarray}
   Q^{_(1)}_{ij} = - \frac{1}{2q}
   \left[
      \nabla_i Q^{(1)}_j + \nabla_j Q^{(1)}_i
   \right].
\label{eqgi14}
\end{eqnarray}
The vector perturbations in the three-velocity are
\begin{eqnarray}
   w^i = \frac{u^i}{u^0} = V^{(1)}(\tau) Q^i,
\label{eqgi15}
\end{eqnarray}
and in the energy momentum tensor are
\begin{eqnarray}
&&
   \delta T^0_{\ \,0} = 0,
\label{eqgi16a}\\
&&
   \delta T^0_{\ \,j} =
   \bar{\rho} [V^{(1)}(\tau) - B^{(1)}(\tau)] Q_j,
\label{eqgi16b}\\
&&
   \delta T^i_{\ \,j} = 0.
\label{eqgi16c}
\end{eqnarray}
For vector perturbations, Bardeen (1980) defined the gauge-invariant
combinations:
\begin{eqnarray}
&&
   \Psi = B^{(1)} - \frac{1}{q} {H^{(1)}_T}',
\label{eqgi17a}\\
&&
    V_c = V^{(1)} - B^{(1)}.
\label{eqgi17b}
\end{eqnarray}
The perturbation equation (\ref{eqgi1}) for these vector
gauge-invariant variables are given by Bardeen (1980):
\begin{eqnarray}
&&
   (q^2 - 2K) \Psi = 16\pi G a^2 \bar{\rho} V_c,
\label{eqgi18a}\\
&&
   {V_c}' + \frac{a'}{a} V_c = 0.
\label{eqgi18b}
\end{eqnarray}
We define a new variable $v_{\rm T} = a^{-1} V_c$, which is the
transverse part of the velocity $v^i$. After superposing the modes of
the equation (\ref{eqgi18b}), we obtain
\begin{eqnarray}
&&
   \dot{v}_{\rm T}^i + 2H v_{\rm T}^i = 0,
\label{eqgi19a}\\
&&
   \nabla_i v_{\rm T}^i = 0.
\label{eqgi19b}
\end{eqnarray}

Since the density contrast and the velocity field are irrelevant to
the tensor perturbations, we do not repeat here equations of motion
for tensor perturbations. Complete equations to determine the density
contrast and the velocity field is given by equations
(\ref{eqgi11a})--(\ref{eqgi11c}) and (\ref{eqgi19a}), where velocity
field is decomposed into the longitudinal part and the transverse
part, as $v^i = \nabla^i\psi + v_{\rm T}^i$. The equation
(\ref{eqgi19a}) is immediately integrated to give
\begin{eqnarray}
   v_{\rm T}^i \propto a^{-2},
\label{eqgi20}
\end{eqnarray}
i.e., the transverse part decays with time. This result is the
consequence of the fact that the matter is pressureless.  In summary,
neglecting the decaying transverse mode, the complete equations to
determine the density contrast and the velocity field is given by
equations (\ref{eqgi11a})--(\ref{eq11c}), where the velocity consists
of purely longitudinal part.

\section*{APPENDIX B\\ ORTHONORMAL SET OF LAPLACIAN IN THREE
DIMENSIONAL SPACES WITH CONSTANT CURVATURES}

\setcounter{section}{2}
\setcounter{equation}{0}

In this appendix, we construct the orthonormal set of the Laplacian in
a space with or without spatial curvature and derive the expression of
the correlation function in real space in terms of the power spectrum,
when the spatial curvature is not negligible.

\subsection{Harmonic Functions}

The Laplacian of a function $Q$ in the metric $\gamma_{ij}$ of the
equations (\ref{eq5.1}) and (\ref{eq1}) is given by
\begin{eqnarray}
   \triangle Q = \left\{
   \begin{array}{ll}
      \displaystyle
      \frac{-K}{\sinh^2\chi}
      \left[
         \frac{\partial}{\partial\chi}
         \left(
            \sinh^2\chi\frac{\partial Q}{\partial\chi}
         \right) +
         \frac{1}{\sin\theta}\frac{\partial}{\partial\theta}
         \left(
            \sin\theta\frac{\partial Q}{\partial\theta}
         \right) +
         \frac{1}{\sin^2\theta}\frac{\partial^2 Q}{\partial\phi^2}
      \right],
      & (K < 0) \\
      \displaystyle
      \frac{1}{\chi^2}
      \left[
         \frac{\partial}{\partial\chi}
         \left(
            \chi^2\frac{\partial Q}{\partial\chi}
         \right) +
         \frac{1}{\sin\theta}\frac{\partial}{\partial\theta}
         \left(
            \sin\theta\frac{\partial Q}{\partial\theta}
         \right) +
         \frac{1}{\sin^2\theta}\frac{\partial^2 Q}{\partial\phi^2}
      \right],
      & (K = 0) \\
      \displaystyle
      \frac{K}{\sin^2\chi}
      \left[
         \frac{\partial}{\partial\chi}
         \left(
            \sin^2\chi\frac{\partial Q}{\partial\chi}
         \right) +
         \frac{1}{\sin\theta}\frac{\partial}{\partial\theta}
         \left(
            \sin\theta\frac{\partial Q}{\partial\theta}
         \right) +
         \frac{1}{\sin^2\theta}\frac{\partial^2 Q}{\partial\phi^2}
      \right],
      & (K > 0)
   \end{array}
   \right.
\label{eqa1}
\end{eqnarray}
To evaluate the inverse Laplacian, it is useful to obtain a complete
set of scalar harmonic functions satisfying the Helmholtz equation
(Harrison 1967; Wilson 1983)
\begin{eqnarray}
 (\triangle + q^2)Q = 0.
\label{eqa2}
\end{eqnarray}
If we separate variables, the angular part of the solution is just a
spherical harmonic $Y_l^m(\theta,\phi)$. The radial part $X_l(\chi)$,
associated with $Y_l^m$ satisfies the following radial equation,
\begin{eqnarray}
&&
   \frac{1}{\sinh^2\chi}
   \frac{\partial}{\partial\chi}
   \left(
      \sinh^2\chi\frac{\partial X_l}{\partial\chi}
   \right) +
   \left[
      \nu^2 + 1 -
      \frac{l(l+1)}{\sinh^2\chi}
   \right] X_l = 0, \qquad (K<0)
\label{eqa2.5a}\\
&&
   \frac{1}{\chi^2}
   \frac{\partial}{\partial\chi}
   \left(
      \chi^2\frac{\partial X_l}{\partial\chi}
   \right) +
   \left[
      \nu^2 -
      \frac{l(l+1)}{\chi^2}
   \right] X_l = 0, \qquad\qquad\qquad\quad\;\; (K=0)
\label{eqa2.5b}\\
&&
   \frac{1}{\sin^2\chi}
   \frac{\partial}{\partial\chi}
   \left(
      \sin^2\chi\frac{\partial X_l}{\partial\chi}
   \right) +
   \left[
      \nu^2 - 1 -
      \frac{l(l+1)}{\sin^2\chi}
   \right] X_l = 0, \qquad\quad\; (K>0)
\label{eqa2.5c}
\end{eqnarray}
where we introduce a new variable $\nu$ as follows
\begin{eqnarray}
&&
   q^2 = -K(\nu^2 + 1), \qquad\qquad (K<0)
\label{eqa3a}\\
&&
   q^2 = \nu^2, \qquad\qquad\qquad\qquad\; (K=0)
\label{eqa3b}\\
&& q^2 = K(\nu^2 - 1). \qquad\qquad\quad (K>0)
\label{eqa3c}
\end{eqnarray}
The solutions of radial equations (\ref{eqa2.5a})--(\ref{eqa2.5c})
which are regular at the origin are given by conical functions, Bessel
functions, and toroidal functions for negative, zero, and positive
curvature, respectively (e.g., Harrison 1967; Abbott \& Schaefer
1986):
\begin{eqnarray}
&&
   X^{\rm (-)}_l(\nu,\chi) =
   (-1)^l M^{\rm (-)}_l(\nu)\sqrt{\frac{\pi}{2\sinh\chi}}
   {\cal P}^{-1/2-l}_{-1/2+i\nu}(\cosh\chi)
\nonumber\\
&& \qquad\qquad\quad =
   \frac{\sinh^l\chi}{\nu}
   \frac{d^l}{d(\cosh\chi)^l}
   \left(\frac{\sin\nu\chi}{\sinh\chi}\right),
   \quad \nu^2 \geq 0, \qquad (K<0)
\label{eqa4a}\\
&&
   X^{(0)}_l(\nu,\chi) = (-1)^l \nu^l j_l(\nu\chi) =
   \frac{\chi^l}{\nu} 
   \left(
      \frac{1}{\chi}\frac{\partial}{\partial\chi}
   \right)^l
   \left(
      \frac{\sin\nu\chi}{\chi}
   \right),
   \quad \nu^2 \geq 0, \quad (K=0)
\label{eqa4b}\\
&&
   X^{\rm (+)}_l(\nu,\chi) =
   (-1)^l M^{\rm (+)}_l(\nu)\sqrt{\frac{\pi}{2\sin\chi}}
   P^{-1/2-l}_{-1/2+\nu}(\cos\chi)
\nonumber\\
&& \qquad\qquad\quad =
   \frac{(-1)^l\sin^l\chi}{\nu}
   \frac{d^l}{d(\cos\chi)^l}
   \left(\frac{\sin\nu\chi}{\sin\chi}\right),
   \quad \nu = 2,3,4,\ldots, \quad (K>0)
\label{eqa4c}
\end{eqnarray}
where ${\cal P}_\nu^\mu$ is the associated Legendre function, and
$P_\nu^\mu(x) = e^{i\pi\mu/2}{\cal P}_\nu^\mu(x + i0)$ is the
associated Legendre function on the cut (Magnus, Oberhettinger \& Soni
1966). The spectrum for the space of positive curvature is discrete
because of the condition that the function is periodic, or
single-valued. We introduce notations, $M^{(\mp)}_l(\nu) = (\nu^2 \pm
1^2)(\nu^2 \pm 2^2)\cdots(\nu^2 \pm l^2)$, and $M^{(\pm)}_0 \equiv 1$.
We can see that $X^{\rm (+)}_l(\nu,\chi) = i^l X^{\rm
(-)}_l(-i\nu,i\chi)$ by analytic continuation with a natural choice of
branches. For $\chi \rightarrow 0$, all these functions behave like
$\chi^l$ with appropriate constants, and thus $X_l(\nu,0) =
\delta_{l0}$ (e.g., Harrison 1967). For the variable $\nu$ is
positive integers in closed universe, $X^{\rm (+)}_l$ can also be
represented by Gegenbauer (or ultraspherical) polynomials
$C^\lambda_n(x)$ as
\begin{eqnarray}
   X^{\rm (+)}_l(\nu,\chi) = 
   \frac{(-2)^l l! \sin^l\chi}{\nu} C_{\nu-l-1}^{l+1}(\cos\chi).
\label{eqa4.5}
\end{eqnarray}
We choose normalizations so that $K\rightarrow 0$ limit of either
equation (\ref{eqa4a}) or (\ref{eqa4c}) reduces to equation
(\ref{eqa4b}). In these normalizations, the derivatives and recursion
relations for $X_l$ are particularly simple. In fact, the derivatives
of the radial functions are simply given by
\begin{eqnarray}
&&
   \frac{\partial}{\partial\chi}
   \left(\frac{X^{\rm (-)}_l(\nu,\chi)}{\sinh^l\chi}\right) =
   \frac{X^{\rm (-)}_{l+1}(\nu,\chi)}{\sinh^l\chi},
   \qquad\qquad (K<0)
\label{eqa5a}\\
&&
   \frac{\partial}{\partial\chi}
   \left(\frac{X^{(0)}_l(\nu,\chi)}{\chi^l}\right) =
   \frac{X^{(0)}_{l+1}(\nu,\chi)}{\chi^l},
   \qquad\qquad (K=0)
\label{eqa5b}\\
&&
   \frac{\partial}{\partial\chi}
   \left(\frac{X^{\rm (+)}_l(\nu,\chi)}{\sin^l\chi}\right) =
   \frac{X^{\rm (+)}_{l+1}(\nu,\chi)}{\sin^l\chi},
   \qquad\qquad (K>0)
\label{eqa5c}
\end{eqnarray}
and the recursion relations, which are derived by the fact that $X_l$
are the solution of the radial equations
(\ref{eqa2.5a})--(\ref{eqa2.5c}), are
\begin{eqnarray}
&&
   (\nu^2 + l^2) X^{\rm (-)}_{l-1}(\nu,\chi) +
   (2l + 1)\frac{\cosh\chi}{\sinh\chi} X^{\rm (-)}_l(\nu,\chi) +
   X^{\rm (-)}_{l+1}(\nu,\chi) = 0,
   \; (K<0)
\label{eqa6a}\\
&&
   \nu^2 X^{(0)}_{l-1}(\nu,\chi) +
   \frac{2l+1}{\chi} X^{(0)}_l(\nu,\chi) +
   X^{(0)}_{l+1}(\nu,\chi) = 0,
   \qquad\qquad\qquad\quad (K=0)
\label{eqa6b}\\
&&
   (\nu^2 - l^2) X^{\rm (+)}_{l-1}(\nu,\chi) +
   (2l + 1)\frac{\cos\chi}{\sin\chi} X^{\rm (+)}_l(\nu,\chi) +
   X^{\rm (+)}_{l+1}(\nu,\chi) = 0.
   \quad (K>0)
\label{eqa6c}
\end{eqnarray}
In this paper, we need only $l=0,1,2,3,4$ for the evaluation of
redshift distortion.  They are given by
\begin{eqnarray}
&&
   X^{\rm (-)}_0 = \frac{\sin\nu\chi}{\nu\sinh\chi},
\label{eqa7a}\\
&&
   X^{\rm (-)}_1 = 
   \frac{1}{\nu\sinh^2\chi}
   ( - \cosh\chi\sin\nu\chi + \nu\sinh\chi\cos\nu\chi),
\label{eqa8a}\\
&&
   X^{\rm (-)}_2 = 
   \frac{1}{\nu\sinh^3\chi}
   \left\{
      \left[
         3 - (\nu^2 - 2)\sinh^2\chi
      \right]
      \sin\nu\chi - 3\nu\sinh\chi\cosh\chi\cos\nu\chi
   \right\},
\label{eqa9a}\\
&&
   X^{\rm (-)}_3 = 
   \frac{1}{\nu\sinh^4\chi}
   \left\{
      \cosh\chi
      \left[
         - 15 + 6 (\nu^2 - 1) \sinh^2\chi
      \right]
      \sin\nu\chi
   \right.
\nonumber\\
&& \qquad\qquad\qquad\qquad +\,
   \left.
      \nu\sinh\chi
      \left[
         15 - (\nu^2 - 11) \sinh^2\chi
      \right]
      \cos\nu\chi
   \right\},
\label{eqa10a}\\
&&
   X^{\rm (-)}_4 = 
   \frac{1}{\nu\sinh^5\chi}
   \left\{
      \left[
         105 - 15(3\nu^2 - 8) \sinh^2\chi +
         (\nu^4 - 35\nu^2 + 24) \sinh^4\chi
      \right]
      \sin\nu\chi
   \right.
\nonumber\\
&& \qquad\qquad\qquad\qquad -\,
   \left.
      \nu\sinh\chi
      \left[
         105 - 10(\nu^2 - 5) \sinh^2\chi
      \right]
      \cos\nu\chi
   \right\},
\label{eqa11a}
\end{eqnarray}
for $K<0$, and
\begin{eqnarray}
&&
   X^{\rm (0)}_0 = \frac{\sin\nu\chi}{\nu\chi},
\label{eqa7b}\\
&&
   X^{\rm (0)}_1 = 
   \frac{1}{\nu\chi^2}
   ( - \sin\nu\chi + \nu\chi\cos\nu\chi),
\label{eqa8b}\\
&&
   X^{\rm (0)}_2 = 
   \frac{1}{\nu\chi^3}
   \left[
      \left(
         3 - \nu^2\chi^2
      \right)
      \sin\nu\chi - 3\nu\chi\cos\nu\chi
   \right],
\label{eqa9b}\\
&&
   X^{\rm (0)}_3 = 
   \frac{1}{\nu\chi^4}
   \left[
      \left(
         - 15 + 6 \nu^2 \chi^2
      \right)
      \sin\nu\chi +
      \nu\chi
      \left(
         15 - \nu^2 \chi^2
      \right)
      \cos\nu\chi
   \right],
\label{eqa10b}\\
&&
   X^{\rm (0)}_4 = 
   \frac{1}{\nu\chi^5}
   \left[
      \left(
         105 - 45\nu^2 \chi^2 +
         \nu^4 \chi^4
      \right)
      \sin\nu\chi -
      \nu\chi
      \left(
         105 - 10\nu^2 \chi^2
      \right)
      \cos\nu\chi
   \right],
\label{eqa11b}
\end{eqnarray}
for $K=0$, and
\begin{eqnarray}
&&
   X^{\rm (+)}_0 = \frac{\sin\nu\chi}{\nu\sin\chi},
\label{eqa7c}\\
&&
   X^{\rm (+)}_1 = 
   \frac{1}{\nu\sin^2\chi}
   ( - \cos\chi\sin\nu\chi + \nu\sin\chi\cos\nu\chi),
\label{eqa8c}\\
&&
   X^{\rm (+)}_2 = 
   \frac{1}{\nu\sin^3\chi}
   \left\{
      \left[
         3 - (\nu^2 + 2)\sin^2\chi
      \right]
      \sin\nu\chi - 3\nu\sin\chi\cos\chi\cos\nu\chi
   \right\},
\label{eqa9c}\\
&&
   X^{\rm (+)}_3 = 
   \frac{1}{\nu\sin^4\chi}
   \left\{
      \cosh\chi
      \left[
         - 15 + 6 (\nu^2 + 1) \sin^2\chi
      \right]
      \sin\nu\chi
   \right.
\nonumber\\
&& \qquad\qquad\qquad\qquad +\,
   \left.
      \nu\sin\chi
      \left[
         15 - (\nu^2 + 11) \sin^2\chi
      \right]
      \cos\nu\chi
   \right\},
\label{eqa10c}\\
&&
   X^{\rm (+)}_4 = 
   \frac{1}{\nu\sin^5\chi}
   \left\{
      \left[
         105 - 15(3\nu^2 + 8) \sin^2\chi +
         (\nu^4 + 35\nu^2 + 24) \sin^4\chi
      \right]
      \sin\nu\chi
   \right.
\nonumber\\
&& \qquad\qquad\qquad\qquad -\,
   \left.
      \nu\sin\chi
      \left[
         105 - 10(\nu^2 + 5) \sin^2\chi
      \right]
      \cos\nu\chi
   \right\},
\label{eqa11c}
\end{eqnarray}
for $K>0$.

Although our normalizations in equations (\ref{eqa4a})--(\ref{eqa4c})
have simple relations for derivatives and recursion relations, it is
also convenient to introduce further normalizations as
\begin{eqnarray}
   \widehat{X}^{(\mp)}_l(\nu,\chi) =
   \frac{X^{(\mp)}_l(\nu,\chi)}{\sqrt{M^{(\mp)}(\nu)}},
   \qquad
   \widehat{X}^{(0)}_l(\nu,\chi) = 
   \frac{X^{(0)}_l(\nu,\chi)}{\nu^l} = (-1)^l j_l(\nu\chi)
\label{eqa12}
\end{eqnarray}
{}From the orthogonality and completeness of conical functions, Bessel
functions, and Gegenbauer polynomials (Magnus et al. 1966), we can
find
\begin{eqnarray}
&&
   4\pi \int \sinh^2\chi d\chi
   \widehat{X}^{\rm (-)}_l(\nu,\chi)
   \widehat{X}^{\rm (-)}_l(\nu',\chi) =
   \frac{2\pi^2}{\nu^2} \delta(\nu - \nu'),
   \qquad (K<0)
\label{eqa12a}\\
&&
   4\pi \int \chi^2 d\chi
   \widehat{X}^{(0)}_l(\nu,\chi)
   \widehat{X}^{(0)}_l(\nu',\chi) =
   \frac{2\pi^2}{\nu^2} \delta(\nu - \nu'),
   \qquad\qquad\; (K=0)
\label{eqa12b}\\
&&
   4\pi \int \sin^2\chi d\chi
   \widehat{X}^{\rm (+)}_l(\nu,\chi)
   \widehat{X}^{\rm (+)}_l(\nu',\chi) =
   \frac{2\pi^2}{\nu^2} \delta_{\nu\nu'},
   \qquad\qquad\quad (K>0)
\label{eqa12c}
\end{eqnarray}
and
\begin{eqnarray}
&&
   \int \frac{\nu^2 d\nu}{2\pi^2}
   \widehat{X}^{\rm (-)}_l(\nu,\chi)
   \widehat{X}^{\rm (-)}_l(\nu,\chi') =
   \frac{\delta(\chi - \chi')}{4\pi\sinh^2\chi},
   \qquad\quad\; (K<0)
\label{eqa13a}\\
&&
   \int \frac{\nu^2 d\nu}{2\pi^2}
   \widehat{X}^{(0)}_l(\nu,\chi)
   \widehat{X}^{(0)}_l(\nu,\chi') =
   \frac{\delta(\chi - \chi')}{4\pi\chi^2},
   \qquad\qquad\; (K=0)
\label{eqa13b}\\
&& \sum_{\nu=2}^\infty \frac{\nu^2}{2\pi^2}
   \widehat{X}^{\rm (+)}_l(\nu,\chi) \widehat{X}^{\rm (+)}_l(\nu,\chi') =
   \frac{\delta(\chi - \chi')}{4\pi\sin^2\chi}. \qquad\qquad (K>0)
\label{eqa13c}
\end{eqnarray}

\subsection{The Correlation Function in Real Space at Present Time}

The density contrast $\delta(\chi,\theta,\phi)$ is expanded in terms
of the eigenfunctions as follows:
\begin{eqnarray}
&&
   \delta(\chi,\theta,\phi) = 
   \sum_{l=0}^\infty \sum_{m=-l}^l
   \int \frac{\nu^2 d\nu}{2\pi^2}
   \tdel_{lm}(\nu)
   \widehat{X}^{\rm (-)}_l(\nu,\chi) Y_l^m(\theta,\phi),
   \qquad(K<0)
\label{eqa14a}\\
&&
   \delta(\chi,\theta,\phi) = 
   \sum_{l=0}^\infty \sum_{m=-l}^l
   \int \frac{\nu^2 d\nu}{2\pi^2}
   \tdel_{lm}(\nu)
   \widehat{X}^{(0)}_l(\nu,\chi) Y_l^m(\theta,\phi),
   \qquad(K=0)
\label{eqa14b}\\
&&
   \delta(\chi,\theta,\phi) = 
   \sum_{l=0}^\infty \sum_{m=-l}^l
   \sum_{\nu=3}^\infty \frac{\nu^2}{2\pi^2}
   \tdel_{lm}(\nu)
   \widehat{X}^{\rm (+)}_l(\nu,\chi) Y_l^m(\theta,\phi).
   \qquad(K>0)
\label{eqa14c}
\end{eqnarray}
For a positive curvature model, the term of $\nu=2$ is omitted because
$\nu=2$ corresponds to a mode which is a pure gauge term (Lifshitz \&
Khalatnikov 1963; Bardeen 1980) as indicated by equation
(\ref{eqgi9a}). The inverse of these expansion is
\begin{eqnarray}
&&
   \tdel_{lm}(\nu) = 
   4\pi \int \sinh^2\chi d\chi
   \int \sin\theta d\theta d\phi\,
   \delta(\chi,\theta,\phi) 
   \widehat{X}^{\rm (-)}_l(\nu,\chi)
   Y_l^{m\,*}(\theta,\phi),
   \; (K<0)
\label{eqa15a}\\
&&
   \tdel_{lm}(\nu) = 
   4\pi \int \chi^2 d\chi
   \int \sin\theta d\theta d\phi\,
   \delta(\chi,\theta,\phi) 
   \widehat{X}^{(0)}_l(\nu,\chi)
   Y_l^{m\,*}(\theta,\phi),
   \qquad\; (K=0)
\label{eqa15b}\\
&&
   \tdel_{lm}(\nu) = 
   4\pi \int \sin^2\chi d\chi
   \int \sin\theta d\theta d\phi\,
   \delta(\chi,\theta,\phi) 
   \widehat{X}^{\rm (+)}_l(\nu,\chi)
   Y_l^{m\,*}(\theta,\phi).
   \quad (K>0)
\label{eqa15c}
\end{eqnarray}

Since the correlation function $\xi(\chi)$ depends only on the
separation of two points, and does not depend on the particular choice
of the coordinate system, we can write it down as
\begin{eqnarray}
&&
   \xi(\chi) =
   \langle \delta(0,\theta,\phi) \delta(\chi,\theta,\phi) \rangle
\nonumber\\
&& \quad =
   \sum_{l,m}
   \sum_{l',m'}
   \int \frac{\nu^2 d\nu}{2\pi^2}
   \int \frac{\nu'^2 d\nu'}{2\pi^2}
\nonumber\\
&& \qquad\,\times
   \widehat{X}^{\rm (-,0)}_l(\nu,0)
   \widehat{X}^{\rm (-,0)}_{l'}(\nu',\chi)
   Y_l^{m\,*}(\theta,\phi) Y_{l'}^{m'}(\theta,\phi)
   \langle \tdel^{\,*}_{lm}(\nu) \tdel_{l'm'}(\nu') \rangle,
   \; (K \leq 0)
\label{eqa16ab}\\
&&
   \xi(\chi) =
   \langle \delta(0,\theta,\phi) \delta(\chi,\theta,\phi) \rangle
\nonumber\\
&& \quad =
   \sum_{l,m}
   \sum_{l',m'}
   \sum_{\nu=3}^\infty \frac{\nu^2}{2\pi^2}
   \sum_{\nu'=3}^\infty \frac{\nu'^2}{2\pi^2}
\nonumber\\
&& \qquad\,\times
   \widehat{X}^{\rm (+)}_l(\nu,0)
   \widehat{X}^{\rm (+)}_{l'}(\nu',\chi)
   Y_l^{m\,*}(\theta,\phi) Y_{l'}^{m'}(\theta,\phi)
   \langle \tdel^{\,*}_{lm}(\nu) \tdel_{l'm'}(\nu') \rangle.
   \quad\; (K>0)
\label{eqa16c}
\end{eqnarray}
Statistical homogeneity and isotropy of the universe suggest that
correlation of modes has the form
\begin{eqnarray}
&&
   \langle \tdel^{\,*}_{lm}(\nu) \tdel_{l'm'}(\nu') \rangle =
   (2\pi)^3 \delta_{ll'} \delta_{mm'}
   \frac{\delta(\nu - \nu')}{\nu^2}
   S(\nu),
   \qquad(K\leq 0)
\label{eqa17ab}\\
&&
   \langle \tdel^{\,*}_{lm}(\nu) \tdel_{l'm'}(\nu') \rangle =
   (2\pi)^3 \delta_{ll'} \delta_{mm'}
   \frac{\delta_{\nu\nu'}}{\nu^2}
   S(\nu),
   \qquad\qquad\; (K>0)
\label{eqa17c}
\end{eqnarray}
where $S(\nu)$ is a power spectrum of the number density field at
present. From this equation and $X_l(\nu, 0) = \delta_{l0}$, equations
(\ref{eqa16ab}) and (\ref{eqa16c}) reduce to
\begin{eqnarray}
&&
   \xi(\chi) = 
   \int\frac{\nu^2d\nu}{2\pi^2}
   X^{\rm (-,0)}_0(\nu,\chi) S(\nu),
   \qquad\qquad (K\leq 0)
\label{eqa18ab}\\
&&
   \xi(\chi) = 
   \sum_{\nu=3}^\infty \frac{\nu^2}{2\pi^2}
   X^{\rm (+)}_0(\nu,\chi) S(\nu).
   \qquad\qquad\quad (K>0)
\label{eqa18c}
\end{eqnarray}
The equation (\ref{eqa18ab}) is an equivalent formula that Wilson (1983)
has previously derived. These formulas can be even transformed to the
forms more like the usual formula:
\begin{eqnarray}
&&
   \xi(x) = 
   \int\frac{k^2dk}{2\pi^2} P(k)
   \frac{\sqrt{-K}\sin(kx)}{k \sinh\left(\sqrt{-K}x\right)},
   \qquad (K<0)
\label{eqa19a}\\
&&
   \xi(x) = 
   \int\frac{k^2dk}{2\pi^2} P(k)
   \frac{\sin(kx)}{kx},
   \qquad\qquad\qquad (K=0)
\label{eqa19b}\\
&&
   \xi(x) =
   \sum_{i = 1}^\infty
   \frac{k_i^{\,2}\sqrt{K}}{2\pi^2} P(k_i)
   \frac{\sqrt{K}\sin(k_i x)}{k_i \sin\left(\sqrt{K}x\right)},
   \qquad (K>0)
\label{eqa19c}
\end{eqnarray}
where the discrete wave number, $k_i = (i+2)\sqrt{K}$, is for the
space of positive curvature. The relations between usual power spectrum
$P(k)$ and the power spectrum $S(\nu)$ are given by
\begin{eqnarray}
&&
   P(k) = 
   \frac{
      S
      \left(|K|^{-1/2}k\right)
   }{|K|^{3/2}},
   \qquad(K\neq 0)
\label{eqa20ac}\\
&&
   P(k) = S(k),
   \qquad\qquad\qquad (K=0)
\label{eqa20b}
\end{eqnarray}
When the scale of interest is much smaller than the curvature scale,
$x \ll |K|^{-1/2}$, both expressions (\ref{eqa19a}) for negative
curvature and (\ref{eqa19c}) for positive curvature reduce to the
familiar expression (\ref{eqa19b}).

Finally, we comment that comparing the expression $\xi(\chi) = \langle
\delta(\chi_1,\theta_1,\phi_1) \delta(\chi_2,\theta_2,\phi_2) \rangle$
and the expression (\ref{eqa18ab}) or (\ref{eqa18c}) proves the
addition theorem for $\widehat{X}_l = \widehat{X}^{(\pm, 0)}_l$:
\begin{eqnarray}
   \widehat{X}_0(\nu,\chi) =
   \sum_l (2l + 1)
   \widehat{X}_l(\nu,\chi_1) \widehat{X}_l(\nu,\chi_2)
   P_l(\cos\theta).
\label{eqa21}
\end{eqnarray}
For the flat case $K=0$, the above equation is nothing but the
well-known addition theorem of the Bessel function, therefore, the
addition theorem for $K \ne 0$ is the generalizations of it to the
constant curvature space.

{}

\newpage


\end{document}